\documentclass[pre,reprint,longbibliography,superscriptaddress]{revtex4-1}

\usepackage{epsfig,amsmath,amssymb,amsthm}
\usepackage[outdir=./]{epstopdf}

\newcommand{\bra}[1]{\langle #1|}
\newcommand{\ket}[1]{|#1\rangle}
\newcommand{\Exp}[1]{\langle#1\rangle}

\begin{document}
	
\title{Quantum chaos and entanglement in ergodic and non-ergodic systems} 

\author{Angelo Piga}
\affiliation{ICFO -- Institut de Ci\`{e}ncies Fot\`{o}niques, The Barcelona Institute of Science and Technology, 08860 Castelldefels, Spain}
\author{Maciej Lewenstein} 
\affiliation{ICFO -- Institut de Ci\`{e}ncies Fot\`{o}niques, The Barcelona Institute of Science and Technology, 08860 Castelldefels, Spain}
\affiliation{ICREA, Pg.~Lluis Companys 23, ES-08010 Barcelona, Spain}
%\author{Fritz Haake}
%\affiliation{Fachbereich Physik, Universit\"{a}t Duisburg-Essen, Lotharstr. 1, 47048 Duisburg, Germany}
\author{James Q. Quach} 
\email{quach.james@gmail.com}
\affiliation{ICFO -- Institut de Ci\`{e}ncies Fot\`{o}niques, The Barcelona Institute of Science and Technology, 08860 Castelldefels, Spain}
\affiliation{Institute for Photonics and Advanced Sensing and School of Chemistry and Physics, The University of Adelaide, South Australia 5005, Australia}

\begin{abstract}
	We study entanglement entropy (EE) as a signature of quantum chaos in ergodic and non-ergodic systems. In particular we look at the quantum kicked top and kicked rotor as multi-spin systems, and investigate the single spin EE which characterizes bipartite entanglement of this spin with the rest of the system. We study the correspondence of the Kolmogorov-Sinai entropy of the classical kicked systems with the EE of their quantum counterparts. We find that EE is a signature of global chaos in ergodic systems, and local chaos in non-ergodic systems. In particular, we show that EE can be maximised even when systems are highly non-ergodic, when the corresponding classical system is locally chaotic. In contrast, we find evidence that the quantum analogue of Kolmogorov-Arnol'd-Moser (KAM) tori are tori of low entanglement entropy. We conjecture that entanglement should play an important role in any quantum KAM theory.
\end{abstract}

\maketitle

\section{Introduction}
\label{sec:Introduction}

Sketched in the 17th century by Newton and others, the deterministic laws of classical mechanics quickly ran into difficulties. Foremost was the fact that equations of Newtonian gravitation resisted analytic solutions 
for three or more bodies. The struggles of the unsolvability of many classical mechanical equations was further exacerbated when Poincair\'{e} proved that perturbation to known integrable solutions in general 
leads to non-integrability or chaos. This was in contrast to observation, which saw nature as substantially regular, from the periodic movements of planets to the sounds of a piano. 
%\textit{How can regular behaviour arise from non-linear equations?} 
This paradox was resolved in the form of KAM theory~\cite{arnold78} which formally explains the persistence of quasi-periodic behaviour in chaotic systems.

% Quantum chaos on the other hand faces a dichotic paradox: \textit{how can chaotic behaviour arise from linear equations}? 
Chaos in classical physics is characterised by a hypersensitivity of the time evolution 
of the system to even small changes in the initial conditions. Classically this is well understood in terms of the hypersensitive dependence of the phase space trajectories. 
Quantum chaos, in contrast, cannot be defined in the same terms, largely due to the fact that there is no general quantum analogue of classical phase space trajectories. As underlying all classical systems are quantum mechanical ones, the confounding question is: how does chaos arise from quantum systems? This question motivates the search for quantum signatures of chaos.

Approaches to quantum signatures of chaos fall into two categories. One involves investigating quantum variables that distinguish between quantum systems whose classical counterparts are integrable and nonintegrable. 
These approaches  typically look at energy spectra properties~\cite{reichl13,haake13,gutzwiller90,tabor89,zyczkowski1990}. A second class of approaches seeks intrinsic quantum definitions of quantum chaos. 
Examples of these include quantum parallels of the Lyapunov exponents and entropy measures~\cite{ford91,schack96,slomczynski96,schack96,schack96a,zurek94,zurek95,Demkowicz2004}. 
There have also been attempts to develop a quantum analogue of KAM theory~\cite{evans2004towards,hose84,brandino15,geisel86}. In this paper we will look at the linear entanglement entropy (EE) as a signature of quantum chaos.

The connection between EE and chaos was first proposed by Zurek and Paz~\cite{zurek94}. Here they studied a classical inverted harmonic oscillator (an unstable but not properly chaotic system) and conjectured that in the corresponding quantum system, weakly coupled to a high temperature bath, the rate of production of the von Neumann entropy equals the sum of the positive Lyapunov exponents. Importantly, this sum is equivalent to the Kolmogorov-Sinai entropy (KSE)~\cite{pesin77}. Even though the conjecture is not directly generalizable to less trivial systems~\cite{miller98}, Zarum and Sarkar~\cite{zarum98} showed a significant correspondence  between the entropy contours of the phase space of the classical kicked rotator (CKR) and the quantum kicked rotator (QKR) embedded in a dissipative environment. Subsequent to the Zurek-Paz conjecture, and perhaps motivated by it, Furuya, Nemes, and Pellegrino numerically showed that classical chaos could be related to high EE and classical regular dynamics to low EE in the context of the Jaynes-Cummings model~\cite{furuya98}. This result stimulated further studies of the EE as a direct signature of chaos.

More recently bipartite EE as a signature of chaos was studied in the quantum kicked top (QKT) modelled as a multi-spin system~\cite{wang04, ghose08,ruebeck2017,kumari2018orbits,kumari2018untangling}, without the need for an external environment. In comparison with the Zurek and Paz model, one side of the bipartite system would serve analogously as the harmonic oscillator, whilst the rest of the system  as the bath. A significant difference however is that in the Zurek and Paz model, the coupling to the bath was weak, whereas in the later works the coupling to the effective bath is strong. In this strong coupling regime, the system decoheres almost instantaneously. These works found that high EE corresponds to chaos in quasi-ergodic systems. Remarkably, it was experimentally observed in a three superconducting spins system~\cite{neill16}. The correspondence between EE and chaos has been argued to not be universal: Lombardi and Matzkin offer a counter-example in the Rydberg molecule~\cite{lombardi11,lombardi15a}. However, their claims are controversial, and have been questioned by others~\cite{madhok15,ruebeck2017}. Recently, Kumari and Ghose~\cite{kumari2018untangling} propose that the conflict arises because Lombardi and Matzkin work deep in the quantum regime, where the correspondence is known to break down.

The correspondence between chaos and EE can be \textit{intuited} when one considers the linear EE measure
\begin{equation}
	S = 1 - \text{tr} \rho_{(A)}^2,
\label{eq:S}
\end{equation}
where $\rho_{(A)}$ is a reduced density matrix of a bipartition of the Hilbert space $\mathcal{H}=A\otimes B$ ($\rho$ is the density matrix of the whole Hilbert space). When $\rho_{(A)}$ is a maximally mixed state, it explores all states equally, \textit{i.e.} it is ergodic. 
In this case $S$ is maximised (see Appendix~\ref{app:Proof that ergodic system maximise entanglement entropy} for proof). For a maximally mixed state, further bipartition of $\rho$ would still result in maximally mixed state, and hence the $S$ would still be maximized.

In classical systems, one may have chaos even in non-ergodic systems. In the CKR for example, in the presence of KAM tori~\cite{arnold78} the system is far from ergodic, yet local chaos exists. Taking this to the quantum regime, it is not immediately obvious that $S$ can be maximised for an analogous non-ergodic $\rho$. We would like to ask: \textit{can bipartite EE be a signature of quantum chaos in highly non-ergodic systems?}

In the present work we tackle this problem by studying in detail both the top and rotor in the classical and quantum regimes. These prototypical systems have the main advantage that they exhibit the most important features of chaotic dynamics and a rich phase space, despite their relative simplicity. In order to directly compare the QKT and the QKR, we study the latter as a special limit of the former, exploiting the formulation given by Haake and Shepelyansky in~\cite{haake88}. In light of the experimental accessibility of the multi-spin system, this approach has the further convenience that it allows one to describe the QKT and QKR in the same closed multi-spin system. Using this system as a case-study, we will show that EE is a signature of quantum chaos even in highly non-ergodic systems. Specifically, in Sec.~\ref{sec:Quantum kicked top} we 
review the CKT and QKT. We calculate the KSE of the CKT and the EE of the QKT. In Sec.~\ref{sec:Quantum kicked rotor} we look at the CKR, the QKR and we derive the kicked rotor as a limiting case of the kicked top. 
We show that bipartite EE is a good signature of quantum chaos in  this non-ergodic system. 
In Sec.~\ref{sec:Quantum Kolmogorov-Arnol'd-Moser Theory} we discuss properties of our quantum system that are reminiscent of KAM theory.

\section{Quantum kicked top}
\label{sec:Quantum kicked top}

The Hamiltonian of the QKT ~\cite{haake86} is
\begin{equation}
	H_T=\alpha J_x + \frac{\beta}{2 j}J_z^2\sum_{n=-\infty}^{\infty}\delta(t-n)
\label{eq:H_T}
\end{equation} 
where $\boldsymbol{J}$ is the angular momentum vector that obeys the commutation relations
\begin{equation}
	[J_i,J_j] = i \varepsilon_{ijk}J_k~.
\label{eq:J_commutation}
\end{equation} 

The magnitude $\boldsymbol{J}^2 = j(j+1)\hbar^2$ is a conserved quantity. The first term in Eq.~(\ref{eq:H_T}) describes a precession around the $x$-axis with angular frequency $\alpha$. 
The second term represents a periodic kick. Each kick is an impulsive rotation around the $z$-axis by an angle proportional to $J_z$. 
For convenience we work in natural units where $\hbar=1$, whereas the time is counted with the number of kicks.
The proportionality factor involves dimensionless coupling constant $\beta/j$, where $\beta$ is known as the torsion strength.

The angular momentum operators at each kick can be obtained from the discrete time evolution of the operators in the Heisenberg picture,
\begin{equation}
	\boldsymbol{J}_{n+1} = U_T^\dagger \boldsymbol{J}_{n} U_T,
\label{eq:heisenbergTop}
\end{equation}
where $U_T$ is the Floquet operator describing the unitary evolution from kick to kick,
\begin{equation}
	U_T = \exp(-i\frac{\beta}{2j} J_z^2)\exp(-i\alpha J_x)~.
\end{equation}

Modelling the system as a $N$-spin system, the angular momentum operators can be expressed in terms of Pauli operators,
\begin{equation}
J_\gamma=\sum_{i=1}^{N}\frac{\sigma_{\gamma_i}}{2}
\end{equation}
where $\gamma = x,y,z$.

We choose the initial pure state to be symmetric under the exchange of any spin, so that the state vector at an later time is also symmetric. 
Thus we can write the state of our $N$-spin system in terms of Dicke states $\ket{j,m}$, where $m=-j,-j+1,\cdots,j$, with $j=N/2$. To connect the quantum and classical dynamics of the kicked top, 
we choose the initial state to be the spin coherent state
\begin{equation}
	\ket{\Phi,\Theta}=\exp\{i\Theta[J_x\sin\Phi-J_y\cos\Phi]\}\ket{j,j}~.
\end{equation}

The state of the system after $n+1$ kick is
\begin{equation}
	\ket{\psi}_{n+1} = U_T\ket{\psi}_n
\end{equation}
where $\ket{\psi}_0=\ket{\Phi,\Theta}$.

\subsection{Classical kicked top}
\label{sec:Classica kicked top}
\begin{figure*}
	\centering
	\includegraphics[width=1.5\columnwidth]{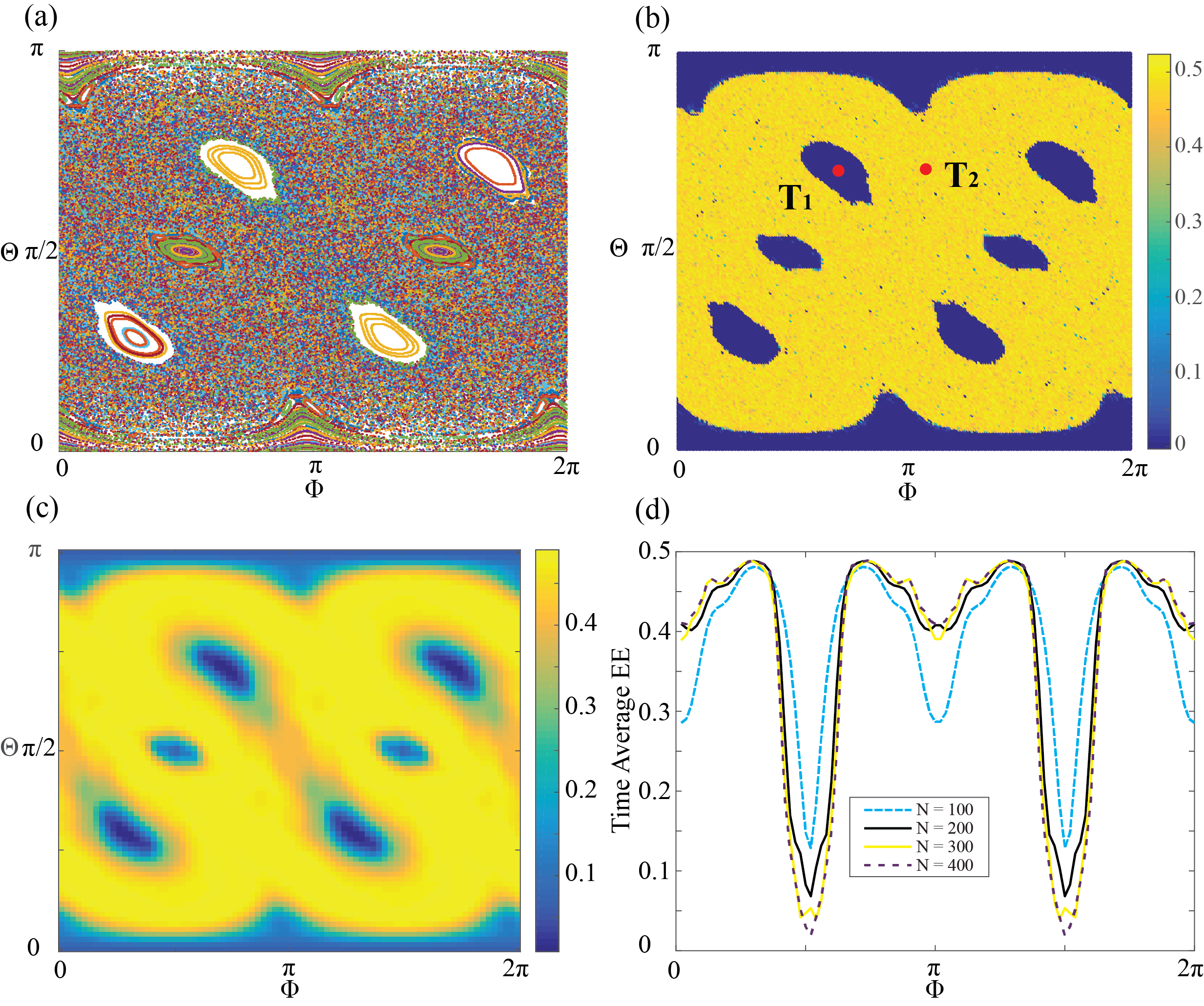}
	\caption{(Color online). (a) The classical phase space of the CKT, with 500 random initial conditions for a duration of 500 kicks. (b) The KSE of the CKT, calculated on a grid of $200\times 200$ initial conditions, iterating the linear map for $10^4$ steps. $\text{KSE} > 0$ corresponds to chaotic behaviour, whereas KSE = 0 indicates regular behaviour. Point \textbf{$T_1, T_2$} marks $(\Phi,\Theta) =(2.20,2.25)$ and $(3.57,2.25)$ respectively. (c) The time-averaged EE of the QKT, calculating for a system of $N=300$ spins and averaged over $T = 300$ kicks. A comparison of (b) and (c) shows a remarkable correspondence between chaotic (regular) classical behaviour and high (low) EE. However, in the classical case there is a well defined demarcation between chaotic and regular regions, whereas in the quantum case the transition from regions of low to high EE is smooth. 
	(d) plots the time-averaged EE at $\Theta=\pi/2$ for different numbers of spins $N$. 
	The transition from regions of low EE to high EE becomes more stark with increasing number of spins, marking the transition to quantum chaos more abruptly, in a similar fashion to classical behaviour. Parameters: $\alpha = \pi/2, \beta = 3$.}
	\label{fig:Top}
\end{figure*}

The classical phase space of the kicked top can be given in the form of a Poincar\'e map, representing the stroboscopic evolution of the classical angular momentum $\boldsymbol{X}_{n+1} = f(\boldsymbol{X}_n)$. As well known in the literature (see e.g.~\cite{haake86}), one can obtain this classical map from the quantum evolution in Eq.~(\ref{eq:heisenbergTop}). The classical angular momentum vector can be parameterised in polar coordinates, $\boldsymbol{X} = (\sin\Theta \cos\Phi,\sin\Theta \sin\Phi,\cos\Theta)$, to give a two-dimensional classical phase space. For the sake of readability we refer to the Appendix~\ref{app:Derivation CKT map} for the explicit form of the map. Fig.~\ref{fig:Top}(a) maps the Poincar\'e map for $\alpha = \pi/2, \beta = 3$. 

\subsection{Top Kolmogorov-Sinai entropy}
\label{sec:Top Kolmogorov-Sinai entropy}

The Poincar\'e map provides a pictorial representation of the phase space, through which one can visually distinguish between regular and chaotic regions. However, to have a proper quantitative measure of the degree of chaoticity, we make use of the KSE. 
The KSE is the rate of change with time of the coarse-grained Gibbs entropy~\cite{kolmogorov58,kolmogorov59} and is calculated as~\cite{pesin77}
\begin{equation}
	h_{\text{KS}} = \lim_{t \to \infty}\frac{1}{t}\sum_{n=1}^{t}\log_2l_n
\label{eq:hKS}
\end{equation}
where $l_n = \sqrt{(\delta X_n)^2 + (\delta Y_n)^2 + (\delta Z_n)^2}$ is the distance in the phase space between two initially close points after $n$ kicks. 
Importantly, Pesin~\cite{pesin77} showed that the KSE is equal to the sum of the positive Lyapunov exponents. 
As the Lyapunov exponents give the rate of separation of two infinitesimally close trajectories, the $\text{KSE} = 0$ for regular regions, and the $\text{KSE} > 0$ for chaotic ones, for times large enough. The KSE therefore is a quantitative measure of the level of chaos. Fig.~\ref{fig:Top}(b) plots $h_\text{KS}$ of the CKT for $\alpha = \pi/2, \beta = 3$. $h_\text{KS} = 0$ for regularly regions. For chaotic regions, $h_\text{KS}>0$, as here the trajectories are divergent. In Appendix~\ref{app:Derivation of tangent space of CKT} a detailed calculation of the KSE is provided.

\subsection{Top Ergodicity}
\label{sec:Ergodicity}

Ergodic systems uniformly explore all states over time, such that observables $O$ averaged over time equals the same observables averaged over all states,
\begin{equation}
\Exp{O}_\text{time} = \Exp{O}_\text{states}~.
\end{equation}

The full QKT system is pure, therefore its EE, defined in Eq.~(\ref{eq:S}), is always zero if we don't take any bipartition of the system. This is not so with its ergodicity. A uniform average over states is given by the microcanonical ensemble $\rho_\text{mc}$, 
therefore the degree to which our system is ergodic is its fidelity with $\rho_\text{mc}$,
\begin{equation}
F(\rho,\rho_\text{mc}) = \text{tr}\sqrt{\sqrt{\rho_\text{mc}}\bar{\rho}\sqrt{\rho_\text{mc}}}
\end{equation}
where $\bar{\rho}$ is the time averaged density matrix of the full system. As all states are equally probable in the microcanonical ensemble, $\rho_\text{mc}$ is a unit matrix of the same dimension as $\rho$. The closer $F$ is to 1, the closer our system is to ergodic behaviour, i.e. time averages are equal to state-space averages. It is worth noting that at time $n$, $\rho_n=\ket{\psi}_n\bra{\psi}_n$ is density matrix of a pure state, but here we are taking the density matrix averaged over time $n$, $\bar{\rho}=\sum_i^n\rho_i/n$. 

In the phase space of the CKT of Fig.~\ref{fig:Top}(a), chaotic initial points explore much of the phase space, in comparison to regular initial points which explore a regular narrow band of the phase space; correspondingly chaotic regions are quasi-ergodic, whereas regular regions are not. 
The chaotic regions are quasi-ergodic,  as there are regular regions which are not visited by initial conditions beginning in the chaotic regions.  Does this notion of chaos and ergodicity hold in the quantum case? To answer this question, we calculate the quantum ergodicity at points corresponding to chaotic and regular initial conditions. 

\begin{figure}
	\centering
	\includegraphics[width=.8\columnwidth]{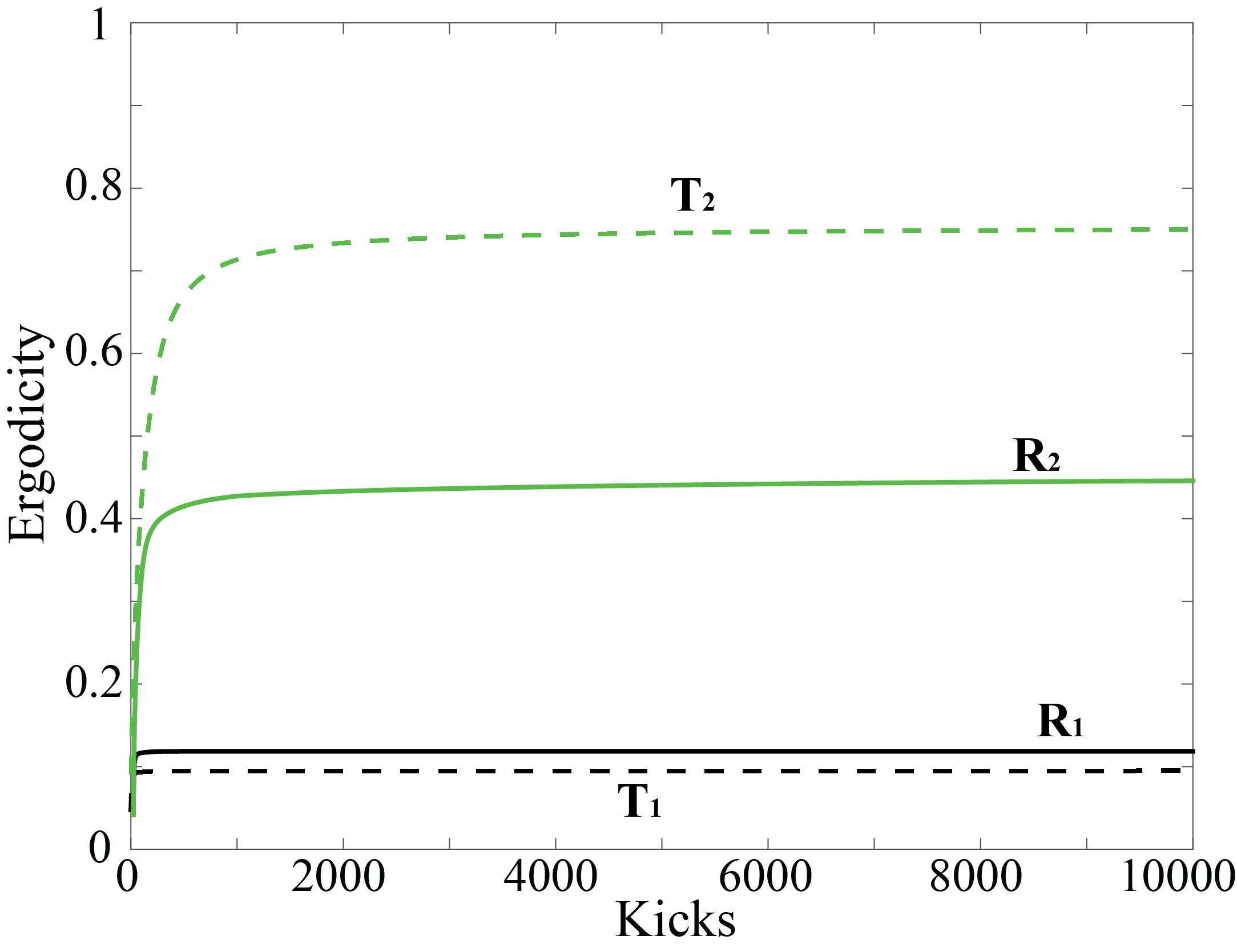}
	\caption{(Color online). Time evolution of the ergodicity for the QKT (dashed lines) and the QKR-limit for $j = 9$ (solid lines) in a regular region (black) and chaotic regions (green/grey). Points $T_1,~T_2$ are marked in Fig.~\ref{fig:Top} and $R_1,~R_2$ are marked in Fig.~\ref{fig:rotor}. The chaotic region in the QKT is quasi-ergodic, but the chaotic region in the QKR-limit is non-ergodic. The number of spins is $N = 500$. }
	\label{fig:ergodicity}
\end{figure}

We pick two representative initial conditions corresponding to regular and chaos points, and calculate their ergodicity in the QKT: $(\Phi,\Theta)=(2.20,2.25)$ and $(3.57,2.25)$, 
labelled respectively as $T_1$ and $T_2$ in Fig.~\ref{fig:Top}. Fig.~\ref{fig:ergodicity} plots the ergodicity of these two points: 
point $T_2$ is quasi-ergodic, whereas point $T_1$ is far from ergodic. In other words, chaotic regions are ergodic and regular regions are not, 
in the corresponding quantum system. In the next section, we calculate the EE of these regions. 

\subsection{Top entanglement entropy}
\label{sec:Top entanglement entropy}

The QKT Hamiltonian acts collectively on all $N$ spins, thereby preserving the symmetry of the $N$-spin state; this means that the spin expectation value of any single spin is
\begin{equation}
	\Exp{s_\gamma} = \frac{\Exp{J_\gamma}}{2j}~.
\end{equation}

The reduced density matrix of a single spin is
\begin{equation}
	\rho_{(1)} = \frac12 + \Exp{\boldsymbol{s}}\cdot\boldsymbol{\sigma}~.
\end{equation}

In the context of quantum chaos and EE, the choice of bipartition is not well understood; different bipartition choice can lead to different results. In prior work, Ref.~\cite{ghose08,lombardi11} bi-partitioned one-particle from the larger systems, Ref.~\cite{wang04} bi-partitioned two-particles from the larger system, and Ref.~\cite{lakshminarayan01} averages over-all possible partitions. However, only the one-particle bi-partition has been experimentally verified~\cite{neill16}. Here we have chosen the one-particle bi-partition, because it is the simplest. The role that the choice of bipartition plays in quantum chaos would make an interesting future study.

From the definition of linear entropy ($S = 1 - \text{tr} \rho_{(1)}^2$), the EE of a single spin with the rest of the system is~\cite{ghose08}
\begin{equation}
	S = \frac12 \Bigg(1-\frac{\Exp{\boldsymbol{J}}\cdot\Exp{\boldsymbol{J}}}{j^2}\Bigg)~.
\label{eq:S}
\end{equation}

We define the time-averaged EE as $1/T \sum_{n} S(n)$, where $S(n)$ is the linear entropy after the $n$ kicks and $T$ is the final number of kicks. For finite systems, time-averaging is used to estimate the equilibrium value approached by larger systems~\cite{neill16}. It is worth noting that the time-average of $S$ is different from applying the time-average of the density matrix to Eq.~(\ref{eq:S}), which  would not be a measure of EE.

We use the time-average EE to depict a quantum phase space: choosing the initial state to be spin coherent $\ket{\Theta,\Phi}$ with $0\le\Theta<\pi$ and $0\le\Phi<2\pi$, Fig.~\ref{fig:Top}(c) plots the time-averaged EE of the QKT for $\alpha = \pi/2, \beta = 3$. Remarkably there is an obvious correspondence between the time-averaged EE of the QKT and the classical phase space trajectories and KSE, as shown in Fig.~\ref{fig:Top}(a) and (b).  Regions of low EE correspond to regular trajectories (KSE = 0), and regions of high EE correspond to chaotic trajectories ($\text{KSE} > 0$). 

An important difference between the KSE and EE of the kicked top however, is that in the classical case there is a well defined demarcation between chaotic and regular regions, whereas in the quantum case the transition from regions of low to high EE is smooth. The change in EE becomes greater with increasing number of spins, and therefore the transition from regions of high EE to low EE occurs more rapidly as shown in Fig.~\ref{fig:Top}(d). From Fig.~\ref{fig:Top}(d) one may conjecture that in the very large spin limit, classical chaotic regions correspond to maximum EE = 1/2, and regular regions corresponds to minimum EE = 0, with a well defined demarcation between these two EE regions, in the QKT.

The surprising correspondence between EE and KSE is made more stark when one compares the vastly different forms of the KSE Eq.~(\ref{eq:hKS}) and the EE Eq.~(\ref{eq:S}). Underlying these very different equations however is a commonality in the information that they encapsulate; both are the rate of information production in their relative classical and quantum domains~\cite{zarum98}.

In the kicked-top, chaotic regions are quasi-ergodic, and these regions are marked by high EE, and non-ergodic regions are marked by low EE. This however is not a general relation, and in the next section we show that non-ergodic regions can also exhibit high EE.

\section{Quantum kicked rotor}
\label{sec:Quantum kicked rotor}

Another well-known kicked system used in the study of chaos is the kicked rotor.  The Hamiltonian of the QKR is
\begin{equation}
	H_R=\frac{1}{2I}P^2+ K \cos\Phi\sum_{n=-\infty}^{\infty}\delta(t-n)
\label{eq:H_R}
\end{equation} 
where $\Phi$ is the angle operator and $P$ is the angular momentum, canonically conjugate to $\Phi$. $K$ is the kicking strength and $I$ is the moment of inertia. 
The rotor operators obey the commutation relation
\begin{equation}
	[P,\Phi] = -i~.
\label{eq:J_communtation}
\end{equation} 

The angular momentum and angle operator at each kick can be obtained from the discrete time evolution of the operators in the Heisenberg picture,
\begin{equation}
\begin{split}
	P_{n+1} &= U_R^\dagger P_n U_R~,\\
	\Phi_{n+1} &= U_R^\dagger \Phi_n U_R~,
\end{split}
\label{eq:heisenbergRotor}
\end{equation}
where the Floquet operator $U_R$ is
\begin{equation}
	U_R = \exp(-i\frac{P^2}{2 I})\exp(-iK\cos\Phi)~.
\end{equation}

This produces the stroboscopic equations
\begin{equation}
\begin{split}
	P_{n+1} &= P_n + K\sin\Phi_n~,\\
	\Phi_{n+1} &= \Phi_n + P_{n+1}/I~.
\end{split}
\label{eq:cRotor}
\end{equation}
As there are no products of $P$ and $\Phi$ terms, this equation is also valid classically.

The phase space of the rotor is a cylinder, $-\infty<P<\infty$, $0\le\Phi<2\pi$. This is topological different than the spherical phase space of the top. 
Although the rotor is unbounded in $P$, the stroboscopic equations show that the system is invariant under $2\pi I$ translations in $P$ and $2\pi$ in $\Phi$. 
Fig.~\ref{fig:rotor}(a) plots the classical rotor phase space for $K=0.9, I=1$.

\subsection{Classical rotor-limit}
\label{sec:Classical rotor-limit}

\begin{figure}
	\centering
	\includegraphics[width=\columnwidth]{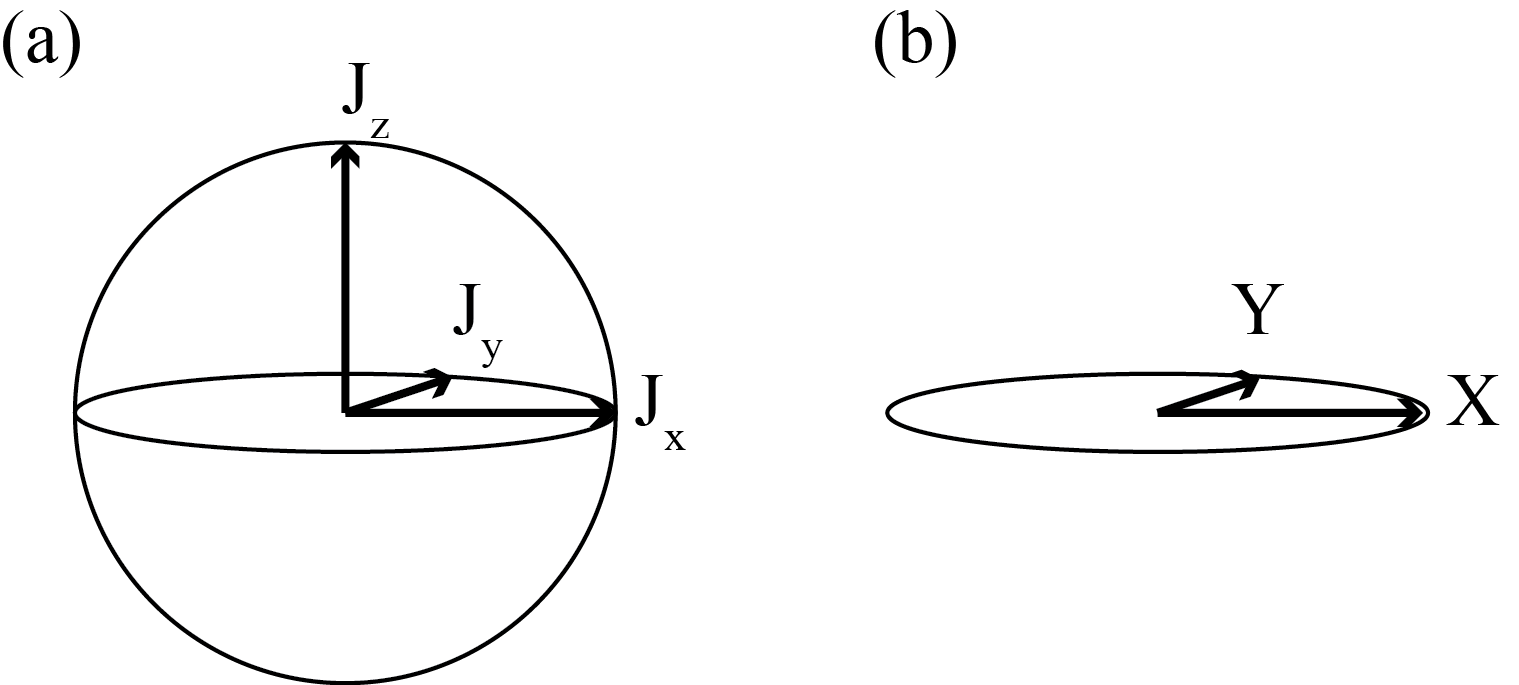}
	\caption{(a) The magnitude of the angular momentum $\boldsymbol{J}$ of the top is a conserved quantity, and therefore is represented on a sphere. (b) The rotor-limit is achieved with the rescaling $\alpha=K/j,\beta=j/I$, where $j\rightarrow\infty$. If one begins in the equatorial waistband, this rescaling confines the angular momentum to $X = \cos\Phi, Y = \sin\Phi, Z = P/j$. }
	\label{fig:rotor_limit}
\end{figure}

Rotor dynamics may by derived from the top if we confine the top to an equatorial waistband as depicted in Fig.~\ref{fig:rotor_limit}~\cite{haake88}. This is achieved by reducing the precession frequency about the $x$-axis and increasing the torsion strength about the $z$-axis through the rescaling
\begin{equation}
	\alpha=K/j~,\quad\beta=j/I~, 
	\label{eq:constraints}
\end{equation}
where $j\rightarrow\infty$. We call this substitution the rotor-limit of the  top, or simply the rotor-limit.

If one begins in the equatorial waistband, this rescaling confines the angular momentum to (Fig.~\ref{fig:rotor_limit})
\begin{equation}
	X = \cos\Phi~,\quad Y = \sin\Phi~,\quad Z = P/j~.
	\label{eq:X_rescaled}
\end{equation} 

Substitution of Eq.~(\ref{eq:constraints}) and (\ref{eq:X_rescaled}) into the kicked-top map of Eq.~(\ref{eq:cTop}), takes one to the kicked-rotor map of Eq.~(\ref{eq:cRotor}). 

\begin{figure*}
	\includegraphics[width=1.5\columnwidth]{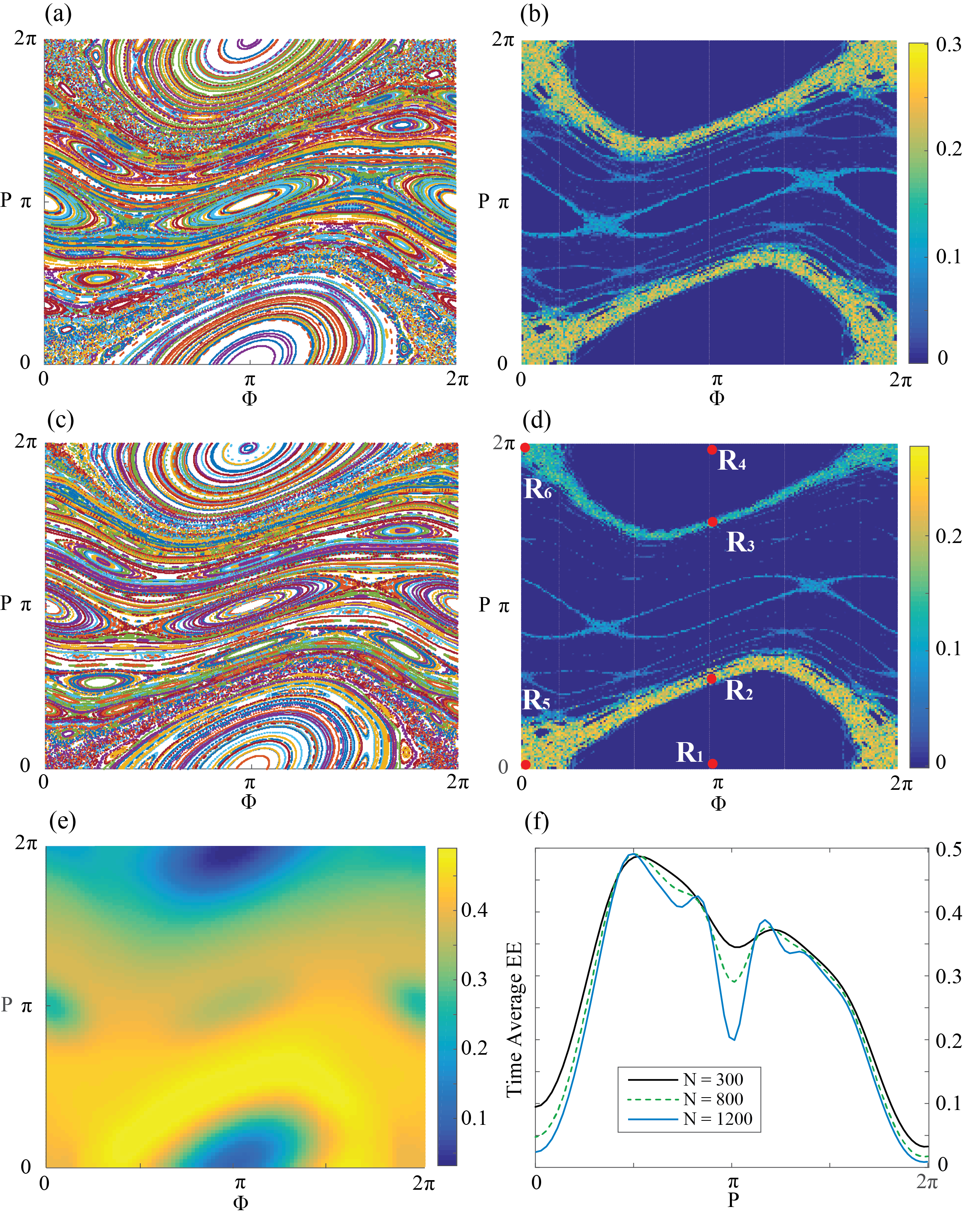}
	\caption{(Color online). (a) and (c) show the classical phase space of the CKR and the CKR-limit ($j=9$), with 500 random initial conditions for a duration of 500 kicks. (b) and (d) show the KSE of the CKR and CKR-limit ($j=9$), calculated on a grid of $200 \times 200$ initial conditions, iterating the linear map for $10^4$ steps. The points \textbf{$R_1,R_2,R_3,R_4,R_5,R_6$} mark $(\Phi,\Theta) = (\pi,0),(\pi,\pi/2),(\pi,3\pi/4),(\pi,2\pi),(0,0),(0,2\pi)$. (e) plots the time-averaged EE of the QKR-limit, for $N = 300$ spins and a duration of $T = 300$ kicks. (f) plots the time-averaged EE of the QKR-limit at $\Phi=\pi$ for various $N$; the transition from regions of low EE to high EE becomes more stark with increasing number of spins. Parameters: $K = 0.9, I = 1$. }
	\label{fig:rotor}
\end{figure*}

The rotor may be approximated by the top even for relatively modest values of $j$. Fig.~\ref{fig:rotor}(c) plots the top phase space with the rescaled $\alpha$ and $\beta$ for $j = 9$, in between $P= 0$ and $2\pi$. A comparison with the rotor phase space of Fig.~\ref{fig:rotor}(a), shows that rotor characteristics are clearly seen in the rotor-limit of the top phase space of Fig.~\ref{fig:rotor}(c). 

\subsection{Rotor Kolmogorov-Sinai entropy}
\label{sec:Rotor Kolmogorov-Sinai entropy}

To further quantity the similarities of the rotor and the rotor-limit, we compare the KSE of the two. From Eq.~(\ref{eq:cRotor}), the tangent map for the rotor is (see Appendix \ref{app:Derivation of tangent space of CKR} for derivation)
\begin{gather}
	\delta P_{n+1} = \delta P_n + K\cos(\Phi_n) \delta \Phi_n~,\\
	\delta \Phi_{n+1} = \Big(1+ \frac{K}{I}\cos\Phi_n\Big) \delta \Phi_n + \frac{\delta P_n}{I}.
\label{eq:tangent_rotor}
\end{gather}

Using this tangent map, Fig.~\ref{fig:rotor}(b) plots $h_\text{KS}$ for the rotor for $K=0.9, I = 1$. Fig.~\ref{fig:rotor}(d) similarly plots  $h_\text{KS}$ for the rotor-limit. The two plots show a high level of similarities, but also differences, which we discuss below.

\subsection{Quantum rotor-limit}
\label{sec:Quantum rotor-limit}

Let us define the following rescaled operators,
\begin{equation}
	\hat{X}\equiv \hat{J}_x/j~,\quad \hat{Y}\equiv \hat{J}_y/j~,\quad \hat{P} \equiv \hat{J}_z~.
\label{eq:J_rescaled}
\end{equation} 

Substitution of these operators into the commutation relations of Eq.~(\ref{eq:J_commutation}), and taking $j\rightarrow\infty$ so that we may drop the $1/j^2$ terms, we get,
\begin{equation}
	[\hat{X},\hat{Y}] = 0,\quad [\hat{Y},\hat{P}] = i\hat{X}~,\quad [\hat{P},\hat{X}] = i\hat{Y}~.
\label{eq:operator_commutations}
\end{equation}

These commutation relations are satisfied with
\begin{equation}
	\hat{X} = \cos\Phi~,\quad \hat{Y} = \sin\Phi~,\quad \hat{P} = -i\frac{\partial}{\partial\Phi}~.
\label{eq:operator_limits}
\end{equation}

Substituting Eq.~(\ref{eq:constraints}), Eq.~(\ref{eq:J_rescaled}), and (\ref{eq:operator_commutations}), into the top Hamiltonian of Eq.~(\ref{eq:H_T}), one retrieves the rotor Hamiltonian of Eq.~(\ref{eq:H_R})~\cite{haake88}.

\subsection{Rotor Ergodicity}

The presence of KAM tori can separate regions of local chaos. This is clearly represented in the phase space of the CKR of Fig.~\ref{fig:rotor}(b), 
where islands of $\text{KSE} > 0$ are separated from each other. The KAM tori act as inpenetrable barriers which prevent the system from exploring the whole phase space; here the system is highly non-ergodic. 
These chaotic regions are localised, as opposed to the global chaos exhibited in the CKT.

We calculate the ergodicity corresponding to a point in one of these regular region and and also in a local chaos region: $(\Theta,\Phi)=(\pi,0)$ and $(\pi,\pi/2)$, respectively. 
These points are marked by $R_1$ and $R_2$ in Fig.~\ref{fig:rotor}(d). Fig.~\ref{fig:ergodicity} shows that the ergodicity of the regular region is low, but that point $R_2$ is also highly non-ergodic. 
We would like to know, whether EE can still be a signature of quantum chaos in these highly non-ergodic regions.

\subsection{Rotor entanglement entropy}
\label{sec:Rotor entanglement entropy}

As the classical rotor may be extracted from the top for modest values of $j$, quantum rotor physics may also be extracted from the quantum top. However in the quantum case we will need a large number of spins to identify the correspondence between EE and chaos in the rotor, as we will show.

We begin with the rotor-limit with $j = 9$, as with the classical example. For the range $0 \le \Exp{\hat{P}} < 2\pi$, $\Theta$ is restricted to $\arccos (2\pi/j) \le \Theta < \pi/2$ or -$\pi/2 \le \Theta < -\arccos (2\pi/j)$, since $P = Zj = j\cos\Theta$. We choose the latter range for $\Theta$, as this will correspond to the rotor map of Eq.~(\ref{eq:cRotor}). For large $j$, this range is a small strip in the equatorial waistband of the top phase space. 

\begin{figure}
	\centering
	\includegraphics[width=\columnwidth]{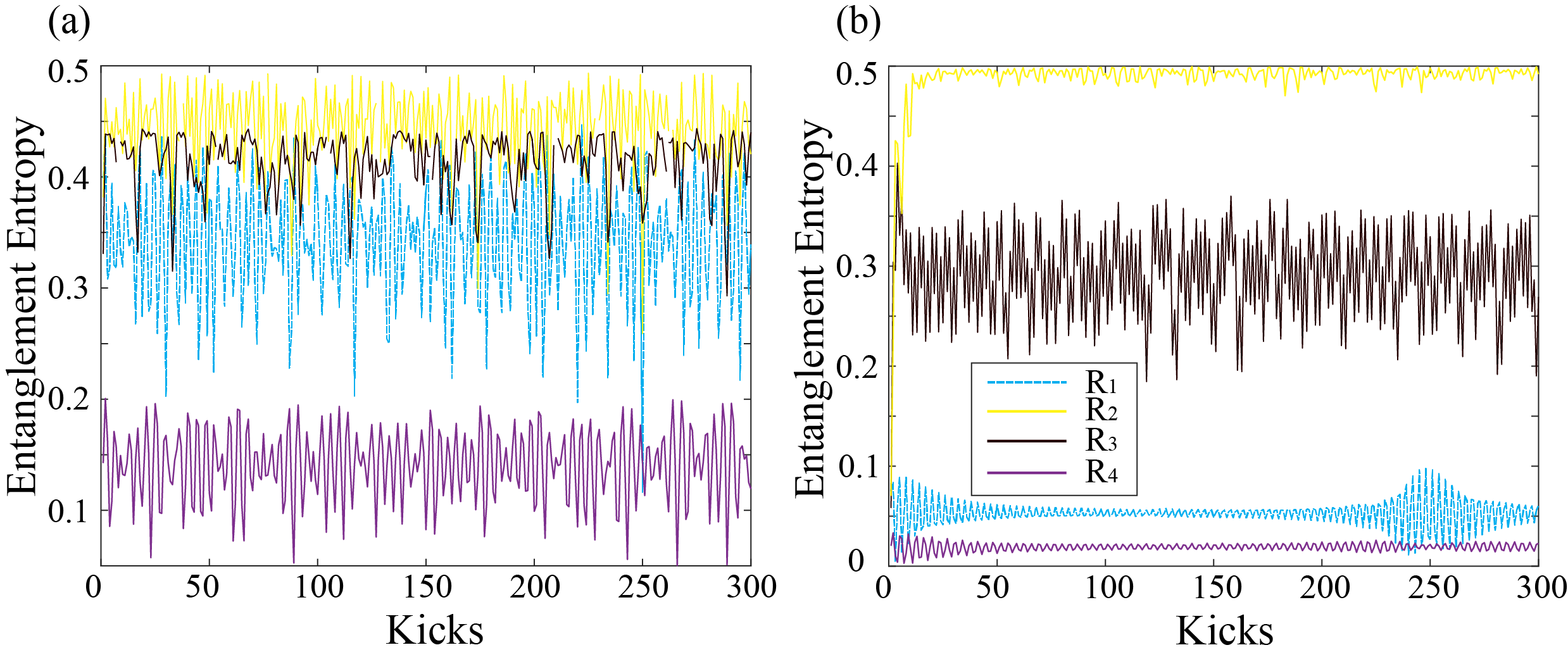}
	\caption{(Color online). Plots of the EE for 4 different points [as marked in Fig.~\ref{fig:rotor}(d)]: $R_1 = (\Phi,P) = (\pi,0)$, $R_2 = (\pi,\pi/2)$, $R_3 = (\pi,3/4\pi)$, $R_4 =(\pi,2\pi)$ for (a) $N = 50$ and (b) $500$. Points $R_1$ and $R_4$ correspond to classical regular behaviour, and points $R_2$ and $R_3$ corresponds to chaos. A comparison of (a) and (b) shows that as one increases the number of spins these regions become more distinguishable, in that there is less overlap of the EE marking each region. Parameters follow Fig.~\ref{fig:rotor}.}
	\label{fig:resolution}
\end{figure}

Now unlike the classical case, where the demarcation between regular and chaotic regions are well defined, in the quantum regime the transition between corresponding regions of low and high EE is gradual. This means that deeper in the quantum regime, features corresponding to the classical features may be washed out. Fig.~\ref{fig:resolution} plots the EE for 4 different points ($K=0.9,I=1$): $R_1 = (\Phi,P) = (\pi,0)$, $R_2 = (\pi,\pi/2)$, $R_3 = (\pi,3/4\pi)$, $R_4 =(\pi,2\pi)$ for $N = 50$ and $500$. From Fig.~\ref{fig:rotor}(b), we see that points $R_1$ and $R_4$ correspond to classical regular behaviour, and points $R_2$ and $R_3$  correspond to chaos. A comparison of Fig~\ref{fig:resolution}(a) and (b) shows that as one increases the number of spins these regions become more distinguishable, in that there is less overlap of the EE marking each region. This is also reflected in Fig.~\ref{fig:rotor}(f) which plots the EE at $\Phi=\pi$ for various $N$, where the difference between EE of chaotic and regular regions becomes greater with increasing $N$.

Fig.~\ref{fig:resolution} shows that points $R_2$ and $R_3$ are different, revealing an asymmetry in the rotor-limit, that is not present in the rotor phase space. This asymmetry is also clearly evident in Fig.~\ref{fig:rotor}(f). In the classical case, the KSE plot of the rotor-limit [Fig.~\ref{fig:rotor}(d)] is also asymmetric, whereas the KSE plot of the rotor is not [Fig.~\ref{fig:rotor}(b)]. The root of the asymmetry lies in the fact that we have used a finite value of $j$, whereas the rotor is reached from the top only in the limit of $j\rightarrow\infty$. 

\begin{figure}
	\centering
	\includegraphics[width=.8\columnwidth]{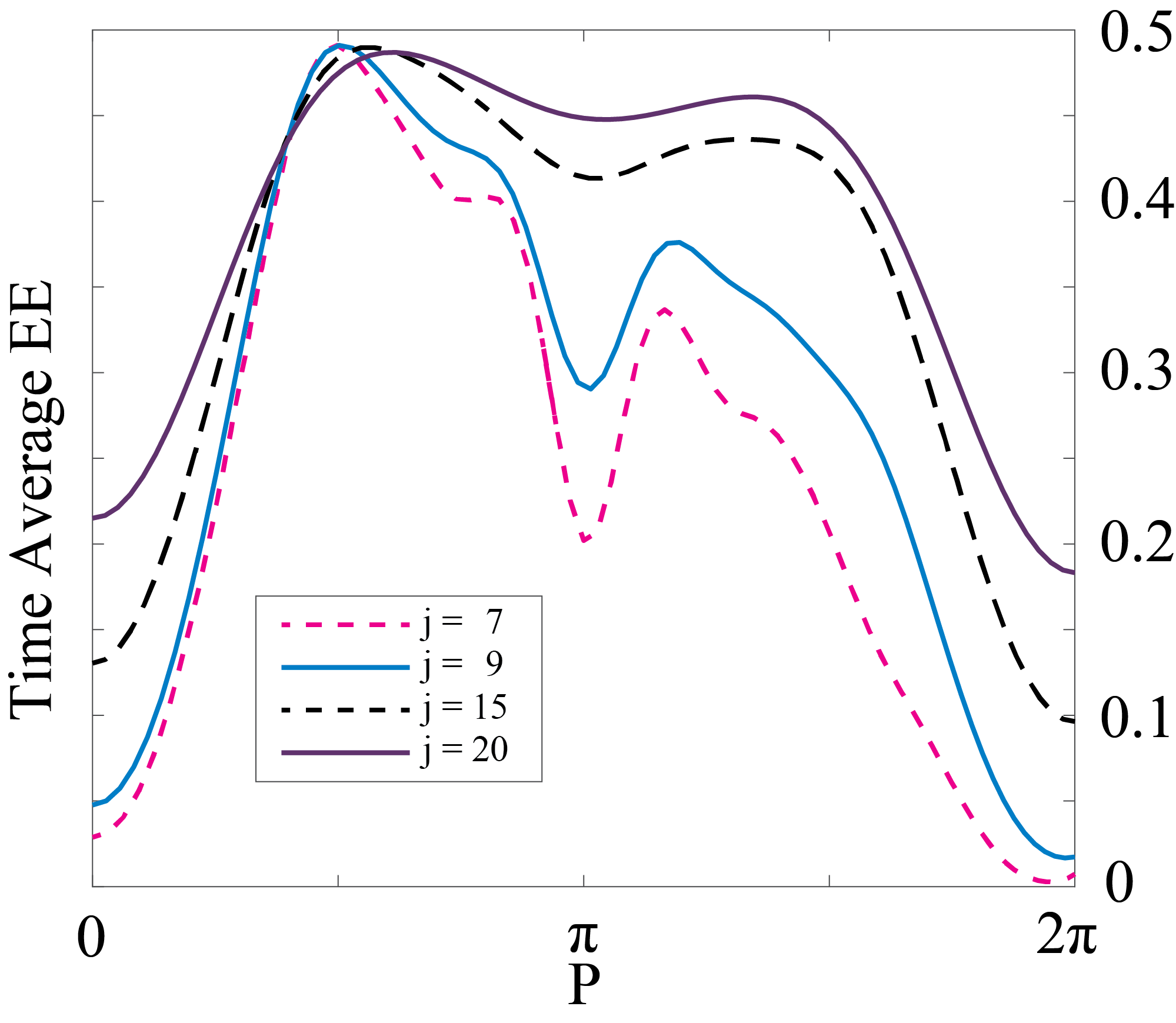}
	\caption{(Color online). The time-averaged EE of the QKT-limit with $N=600$ spins, at $\phi=\pi$ for various $j$. As $j$ increases, the EE is more symmetric as it approaches the QKR. }
	\label{fig:asymmetry}
\end{figure}

Fig.~\ref{fig:asymmetry} plots the time-averaged EE at $\Phi=\pi$ for various $j$ with constant $N=500$. Now two operational properties of the rotor-limit is revealed here. Firstly, increasing $j$ means that the behaviour of the system approaches that of the quantum rotor, thereby reducing the aforementioned asymmetry. Secondly, increasing $j$ for a constant $N$ presents a trade-off: although the system approaches the quantum rotor, for larger values of $j$ one requires more spins to achieve the correspondence between EE and the classical features of the phase space; i.e. the difference between the EE of chaotic and regular regions is reduced. The intuitive reason for this is that, larger $j$ means that we are working in a narrower equatorial waistband. In the parameters of the top, this means that $\Theta$ is confined to the range $\{\pi/2,\arccos(2\pi/j)\}$. As we are working in an increasingly narrower region as $j$ increases, one requires a larger number of spins to be able to distinguish the corresponding classical features, as exemplified in Fig.~\ref{fig:resolution}.

Finally we plot the EE corresponding to the entire classical phase space for the QKR-limit in Fig.~\ref{fig:rotor}(e). We see a qualitative correspondence with the KSE of the CKR-limit in Fig.~\ref{fig:rotor}(d). Importantly, in contrast to the kicked top, here the system is far from ergodic. \textit{Therefore EE can be a signature of quantum chaos even in non-ergodic systems}. 

To intuit how this is so, recall Eq.~(\ref{eq:S}) which gives the EE of the one-spin bi-partition. For maximally mixed states, the state space is explored uniformly, so that $\Exp{\boldsymbol{J}}=0$. One may roughly consider this to be the case for the quasi-ergodic QKT in Fig.~\ref{fig:Top}. However, one should only consider this as an intuitive explanation, as our system is in fact not maximally mixed; it is the time-averaging of procedure that gives rise to these effects. In the rotor-limit however, exploration in the $\boldsymbol{J}_z$ direction is suppressed, so that $\boldsymbol{J}_z\rightarrow 0$, for states beginning in the equatorial waistband. This suppression means that not all state space are uniformly explored, and therefore the system is far from being ergodic. This means that, $\Exp{\boldsymbol{J}}=0$ under the conditions that  $\Exp{\boldsymbol{J}}_x = \Exp{\boldsymbol{J}}_y =0$; under these conditions EE is maximised even though the system is non-ergodic conditions. In the next section, we provide an alternative explanation which is reminiscent of KAM theory.

\section{Quantum Kolmogorov-Arnol'd-Moser Theory}
\label{sec:Quantum Kolmogorov-Arnol'd-Moser Theory}

An integrable Hamiltonian $H_0$ in the presence of perturbation is written as
\begin{equation}
	H = H_0(\boldsymbol{\kappa}) + \epsilon V(\boldsymbol{\kappa},\boldsymbol{\lambda})
\end{equation}
where $\boldsymbol{\kappa}$ and $\boldsymbol{\lambda}$ are the action variables with total $D$ dimension, and $\epsilon$ is a small perturbation parameter. Integrable $H_0$ generates periodic phase-space trajectories that lie on $D$-dimensional tori surfaces. The KAM theorem states that for sufficiently small $\epsilon$, the tori of $H_0$ do not vanish but are deformed, so that the trajectories generated by $H$ are conditionally periodic~\cite{arnold78}. What is the quantum analogue of KAM theory? 
 
Understanding the crossover behaviour arising from the integrability breaking in quantum systems have been pursed through indirect measures such as level statistics~\cite{berry77,brandino10,rigol09,rigol09a} and in the quasi-classical limit of systems using semi-classical eigenfunction hypothesis~\cite{percival73,berry77a,voros79}. In more direct analogy with KAM theory, an existence conditions for localisation in non-integrable quantum systems has also been developed~\cite{hose83}. We do not give a quantum KAM theory here, but simply show properties in our quantum system that are reminiscent of KAM theory. Our motivation is that this may lead to a robust quantum KAM theory in future work.

\subsection{Entanglement entropy and quantum KAM tori}
\label{sec:Entanglement entropy as a signature of quantum KAM tori}

An alternative perspective on why regions with maximum EE may not be ergodic, can be found by considering the CKR in the context of KAM theory~\cite{chirikov79,greene79}. For very small kicking strength trajectories are regular, for very large kicking strength trajectories are chaotic. In between these two extremes, both types of trajectories exists in the phase space, with islands of chaotic regions separated by KAM tori. This means that these islands of chaotic trajectories are bounded and do not explore the whole phase space. The critical value of $K$ where the last of the KAM tori disappears is $K_C \simeq 0.971635$ ($I=1$)~\cite{greene79}. 

\begin{figure}
	\centering
	\includegraphics[width=0.8\columnwidth]{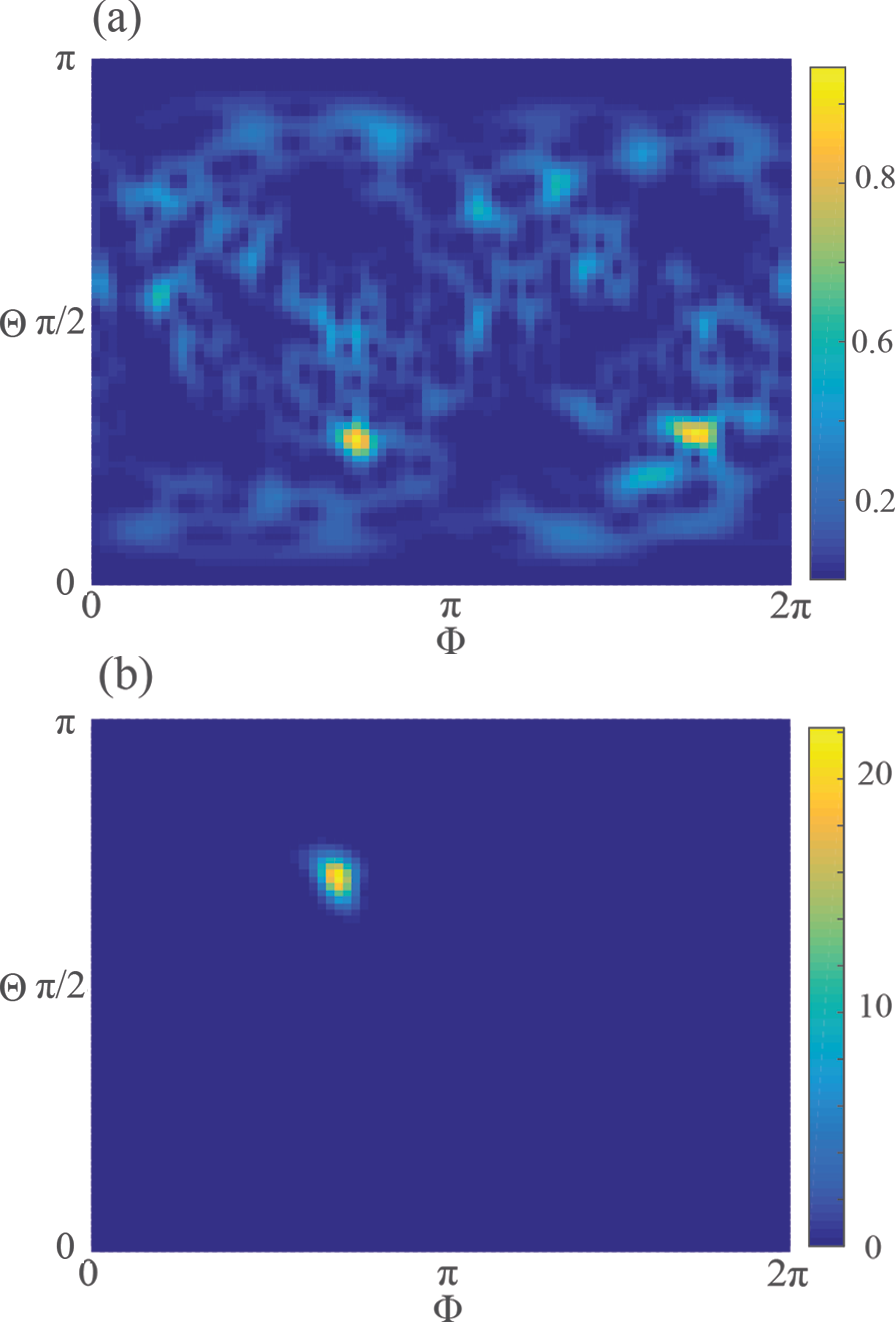}
	\caption{The Husimi distribution for the QKT after 500 kicks. (a) The initial state corresponds to chaotic point $T_2$ in Fig.~\ref{fig:Top}. The resulting Husimi distribution spreads across the phase space.  (b) The initial state corresponds to regular point $T_1$ in Fig.~\ref{fig:Top}. The resulting Husimi distribution is highly confined.  All the parameter of the QKT are the same as in Fig.~\ref{fig:Top}(c).}
	\label{fig:Husimi-top}
\end{figure} 

A quantum analogue of this classical behaviour can be qualitatively inspected by using the Husimi distribution ~\cite{husimi1940,takahashi1985}. At each point $(\Phi,P)$ of the quantum phase space, the Husimi distribution gives the expectation value of the density matrix $\rho_n$ (at time $n$) of the spin coherent state $\ket{\Phi, P}$,
\begin{equation}
\mathcal{P}_H(\Phi,P) =\frac{2j + 1}{4\pi} \bra{\Phi, P} \rho_n \ket{\Phi, P}.
\end{equation}

%It is obtained from a Gaussian smoothing of the Wigner function but, differently from this one, the Husimi distribution is everywhere non negative.
In Fig.~\ref{fig:Husimi-top}, we apply the Husimi distribution to the QKT for states initialised in the regular region (point \textbf{$T_1$}) and chaotic region (point \textbf{$T_2$}). Here the Husimi distribution floods the phase space for states initialised in the chaotic region [Fig.~\ref{fig:Husimi-top}(a)], and is localised for states initialised in a regular region [Fig.~\ref{fig:Husimi-top}(b)]. 

\begin{figure*}
	\includegraphics[width=1.5\columnwidth]{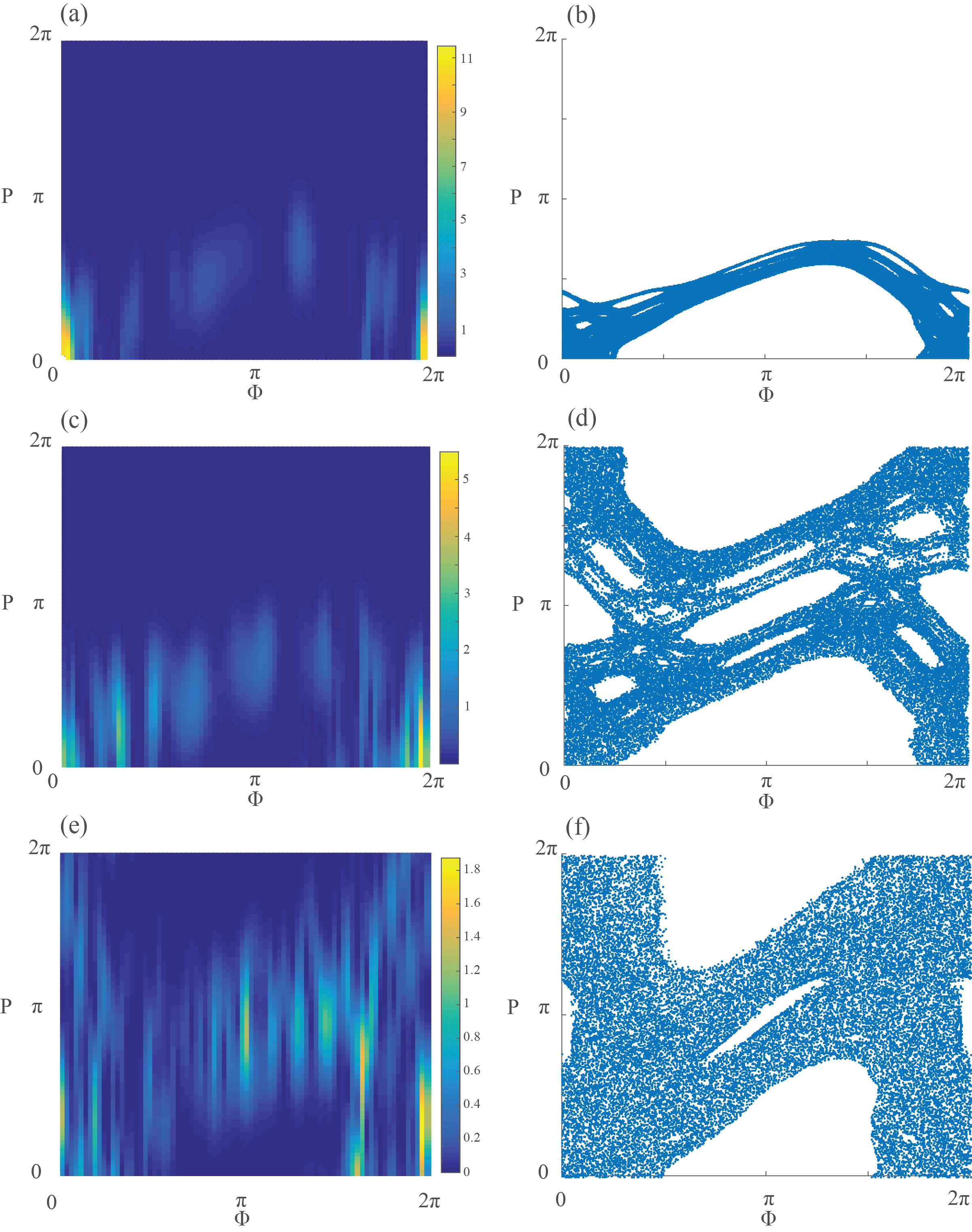}
	\caption{A comparison of the Husimi distributions [(a), (c) and (e)] for the QKR-limit and trajectories for the CKR [(b), (d) and (f)], for the initial state corresponding to point $R_5$ [in Fig.~\ref{fig:rotor}(d)], after 500 kicks, for $N=500$. In (a) and (b) the kicked strength $K = 0.9$ is below the critical point $K_C$, and KAM tori confines the Husimi distribution and the classical trajectories. In (c) and (d), $K=1.2$ is just above $K_C$: in the quantum case the Husimi distribution is confined (c), whereas the classical trejctory spreads across the phases space. Here the KAM cantori is impenetrable in the quantum case, but permeable in the classical case. Finally, in (e) and (f) the kicked strength $K = 2$ is well above the critical point $K_C$, and the absence of KAM tori and cantori means the Husimi distribution and the classical trajectory can flood the phase space.}
	\label{fig:Husimi-rotor}
\end{figure*}

In Fig.~\ref{fig:Husimi-rotor}(a), (c) and (e), we apply the Husimi distribution to the QKR-limit for states initialised in the chaotic region (point \textbf{$R_5$}) for increasing kicked strengths: $K = 0.9$ (below the classical critical point $K_C$), $K = 1.2$ (just above $K_C$) and $K = 2$ (well above $K_C$). We observe that although all the three cases correspond to chaos, it is only when $K$ is well above $K_C$ does the Husimi distribution flood the phase space [Fig.~\ref{fig:Husimi-rotor}(e)]; in comparison, for $K$ near and below $K_C$, the Husimi distribution is confined to a much smaller fraction of the phase space [Fig.~\ref{fig:Husimi-rotor}(a) and (c)]. This is in constrast to the QKT, where chaos corresponds to a Husimi distribution which floods the phase space.

We compare the Husimi distribution of the QKR-limit with the classical trajectories of th CKR-limit. In Fig.~\ref{fig:Husimi-rotor}(b), (c) and (d), we plot the trajectories of the CKR-limit for states initialised in the chaotic region (point \textbf{$R_5$}) for the same values of $K$ as for the QKR-limit. When $K$ is above the critical point the trajectory floods the phase space [Fig.~\ref{fig:Husimi-rotor}(d) and (f)]; in comparison, when $K$ is below the critical point, the trajectory is confined to a much smaller fraction of the phase space [Fig.~\ref{fig:Husimi-rotor}(b)]. Classically, this confinement marks the boundary of the impenetrable regular KAM tori. A comparison of the Husimi distribution of the QKR-lmit with the classical trajectories of the CKR-limit for $K$ well above and below $K_C$ shows a clear qualitative resemblance. This behaviour suggests the presence of a quantum analogue of the KAM tori and a phase transition. An important difference however occurs near $K_C$, which will discuss later in this section. 

\begin{figure*}
	\centering
	\includegraphics[width=1.5\columnwidth]{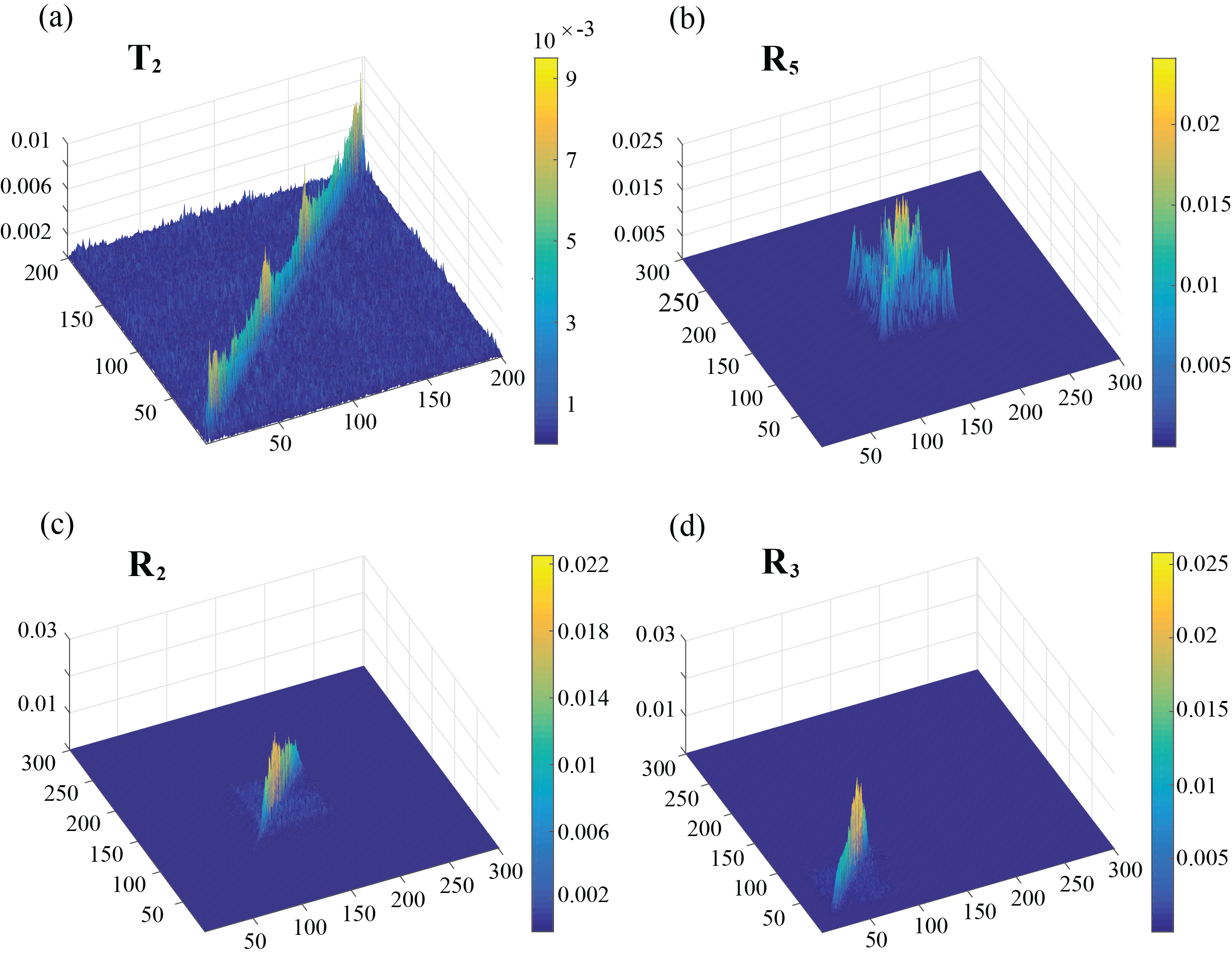}
	\caption{(Color online). (a)-(d) shows the time-averaged density matrices for point \textbf{$T_2$} in Fig.~\ref{fig:Top}(c) and points \textbf{$R_5$},\textbf{$R_2$},\textbf{$R_3$} in Fig.~\ref{fig:rotor}(d), respectively. \textbf{$R_2$} and \textbf{$R_5$} belong to the same island of high EE, 
	whereas \textbf{$R_3$} belongs to a different high EE island. \textbf{$T_1$} uniformly explores the full Hilbert space and therefore is ergodic. \textbf{$R_2$},\textbf{$R_3$},\textbf{$R_5$} 
	explore a subset of the full Hilbert space, and therefore is not ergodic. \textbf{$R_2$} and \textbf{$R_3$} explore different regions of the subspace, whilst \textbf{$R_2$} and \textbf{$R_5$} 
	explore the same subspace, leading to the notion that evolution from initial conditions belonging to the same island will explore the same subspace, but different islands explore different subspaces.}
	\label{fig:density}
\end{figure*}

One may consider torus regions of low EE as the quantum counterpart of classical KAM tori, separating islands of high EE. By analogy to the classical case, we conjecture that the presence of torus regions of low EE indicate that states beginning in different islands of high EE will explore mutually exclusive states, i.e. low EE tori separate orthogonal states. This conjecture is numerically supported in Fig.~\ref{fig:density}(a)-(d) which shows the time-averaged density matrices of the full system with initial conditions marked by \textbf{$T_2$} in Fig.~\ref{fig:Top}(b), and initial conditions marked by \textbf{$R_5$},\textbf{$R_2$},\textbf{$R_3$} in Fig.~\ref{fig:rotor}(b). Fig.~\ref{fig:density} graphically depicts the magnitude of the elements of the time-averaged density matrices. Note that we use larger values of $j=15$ and $N=500$ than that used to generate Fig.~\ref{fig:rotor}(e), to be closer to the rotor-limit. The graphical representation clearly shows that \textbf{$T_2$} explores the full Hilbert space, whilst \textbf{$R_2$},\textbf{$R_3$}, \textbf{$R_5$} only explore a subset of the Hilbert space, explaining why these latter systems are not ergodic.

\textbf{$R_2$} and \textbf{$R_5$} explore the same subspace, whilst \textbf{$R_2$} and \textbf{$R_3$} explore different different subspace. 
Inspection of the rotor phase space in Fig.~\ref{fig:rotor}(b) shows that \textbf{$R_2$} and \textbf{$R_5$} belong to the same chaotic island, whilst \textbf{$R_3$} belongs to a different one, separated by KAM tori. 
This leads to the notion that evolution from initial conditions belonging to the same high EE island will explore the same subspace, but different islands explore different subspaces. It is surrounded by a sea of low EE tori that acts to prevent the subspace overlap between different islands.

Increasing the kicking strength above some critical point destroys the low EE tori, and the system is free to explore the full Hilbert space, as for example represented with point \textbf{$T_2$}; here the system uniformly explores the full Hilbert space and therefore has high fidelity with the microcanonical ensemble. Interestingly the critical point which sees the destruction of low EE tori and the onset of ergodicity, corresponds near to the classical $K_C$ value. In the CKR, $K_C$ marks the disappearance of the last KAM tori and the formation of KAM cantori (broken tori~\cite{mackay84,bensimon84}). 
In the classical case, KAM tori act as impenetrable barriers to the growth of the mean square displacement, whereas KAM cantori are permeable barriers which only slow the diffusive growth. In contrast, Geisel \textit{et al.}~\cite{geisel86} showed that in the quantum case, both KAM tori and cantori correspond to the prohibition of diffusive growth.

\begin{figure}
	\centering
	\includegraphics[width=.8\columnwidth]{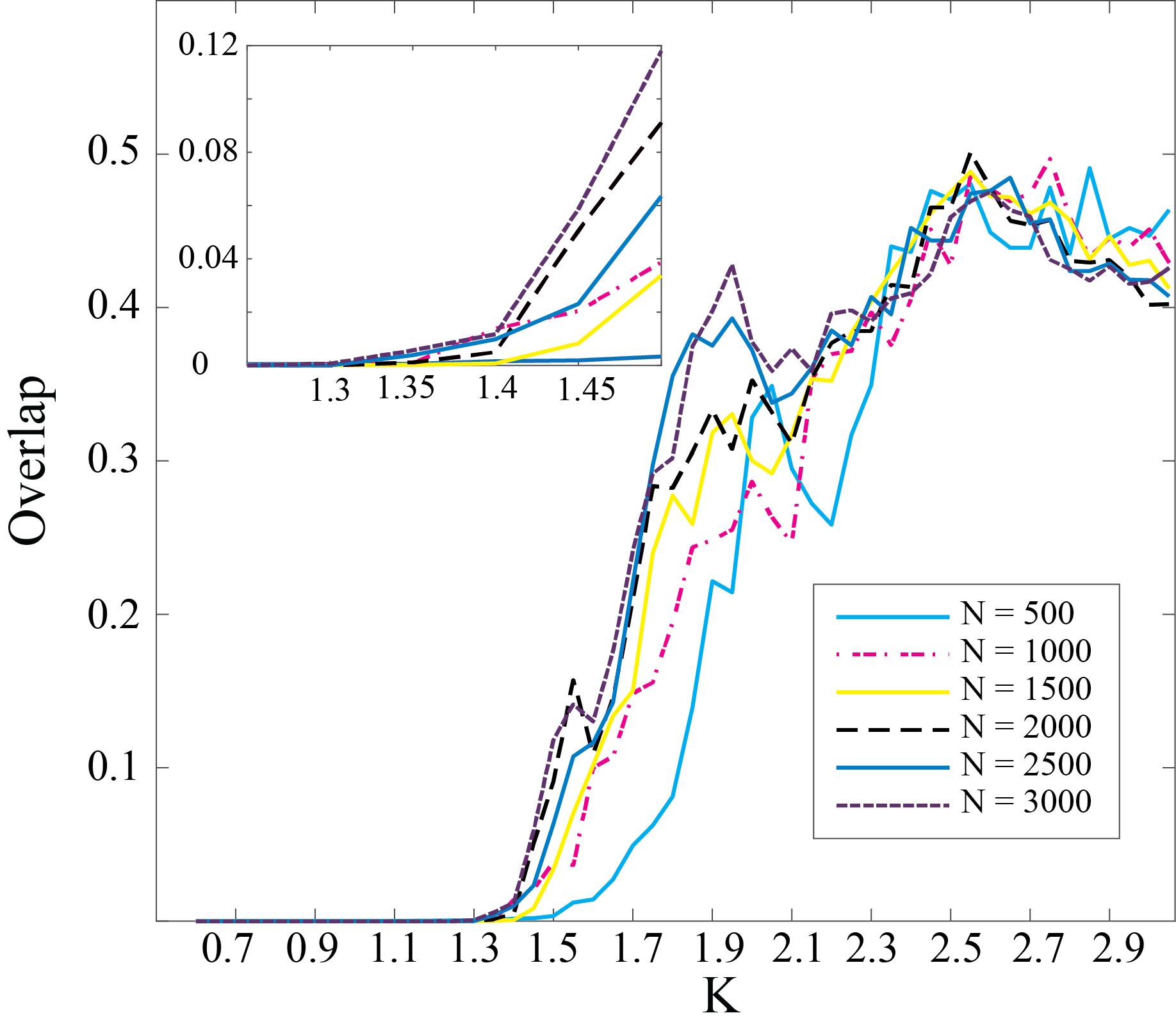}
	\caption{(Color online). The time averaged fidelity of points $R_5=(0,0)$ and $R_6=(0,2\pi)$, $F(\hat{\rho}_{R_5},\hat{\rho}_{R_5})$, 
	as a function of $K$ for $N=500,1500,2000,2500,3000$. The fidelity of points \textbf{$R_5$} and \textbf{$R_6$} begin to increase not at $K_C \simeq 0.971635$, but just after, 
	supporting the idea that cantori correspond to an impenetrable barrier in the quantum case.}
	\label{fig:fidelity}
\end{figure}

To locate the quantum critical point, corresponding to the breaking of the of quantum cantori, in Fig.~\ref{fig:fidelity}, we plot the time averaged fidelity of points \textbf{$R_3$} and \textbf{$R_5$} [$F(\hat{\rho}_{R_3},\hat{\rho}_{R_5})$] as a function of $K$ for $N=500,1000,2000,3000$. 
Remarkably, the fidelity of points \textbf{$R_3$} and \textbf{$R_5$} begin to increase not at $K_C$, but just after, supporting the Giesel \textit{et al.'s} result that cantori correspond to an impenetrable barrier in the quantum case. In other words, Fig.~\ref{fig:fidelity} suggests that between $K=0.97$ and $1.3$, there exists low EE cantori that prevents diffusive growth. This behaviour is supported in the Husimi distribution of Fig.~\ref{fig:Husimi-rotor} (c): in the quantum case the Husimi distribution is localised (c), whereas the classical trajectory has spread across the phase space (d). This is because here, $K$ is not too far above the critical point. Further increases in $K$ sees the Husimi distribution spread across the phase space (e). 

The behaviour of the EE described here are highly reminiscent of the properties of classical KAM tori, supporting our conjecture that the quantum equivalent of KAM trajectories are tori regions of low EE. The Husimi distribution is a useful tool for locating the quantum cantori. These properties suggests that the methods used and quantum systems studied here, are a fruitful avenue to study quantum KAM theory.

\section{Conclusion}
We have explored the correspondence of EE and chaos in the kicked top and kicked rotor. We have shown that high EE corresponds to global chaos in ergodic systems. The key novelty of the present paper relies on the careful use of the rotor-limit of the kicked top.
An important original result is that in taking the rotor-limit of the kicked top, we have also shown that EE corresponds to local chaos in non-ergodic systems. We have shown that the behaviour of EE tori resembles that of KAM tori, and therefore propose that entanglement should play an important role in any quantum KAM theory. Another interesting avenue of future investigation would be to formally understand what the role of bi-partition choice plays in quantum chaos. At the interface of experimental accessibility and theoretical analysis, the QKT and QKR continue to be fruitful systems in which to study quantum chaos. 

\section*{Acknowledgements}
\label{sec:Acknowledgements}
We thank F. Haake for providing feedback on the manuscript.  We acknowledge support of EU Marie Curie Actions for Co-funding of Regional, National and International Programmes (COFUND), the Spanish Ministry MINECO (National Plan 15 Grant: FISICATEAMO No. FIS2016-79508-P, SEVERO OCHOA No. SEV-2015-0522), Fundaci\'o Cellex, Generalitat de Catalunya (AGAUR Grant No. 2017 SGR 1341 and CERCA/Program), ERC AdG OSYRIS, EU FETPRO QUIC, the National Science Centre, Poland-Symfonia Grant No. 2016/20/W/ST4/00314, and the Ramsay Fellowship.

\appendix
\begin{widetext}
\section{Proof that ergodic system maximise entanglement entropy}
\label{app:Proof that ergodic system maximise entanglement entropy}

The linear entanglement entropy defined as $S = 1 - \text{tr} \rho^2$ is maximized when $\text{tr}\rho^2$ is minimized. The density matrix $\rho$, is an Hermitian operator of unitary trace. 
We want to prove that $\text{tr}\rho^2$ is minimised if all the diagonal elements $\rho_{11} =\rho_{11} = \dots = \rho_{NN} = 1/N$, i.e. if the system is ergodic. This is a constrained optimization problem, 
solvable with the Lagrange multipliers method. The constraint is $\text{tr}\rho = 1$. The quantity to minimize is $\text{tr} \rho^2 = \rho_{11}^2 + \rho_{11}^2 +\dots +\rho_{NN}^2$. We define the Lagrangian as:
\begin{equation}
\mathcal{L} = \rho_{11}^2+\rho_{22}^2+\dots+\rho_{NN}^2 - \lambda \big(\rho_{11}+\rho_{22}+\dots+\rho_{NN} -1 \big).
\end{equation}
We want to minimize $\mathcal{L}$ with respect to $\rho_{ii}$ and $\lambda$, $\frac{\partial \mathcal{L}}{\rho_{ii}} = \frac{\partial \mathcal{L}}{\lambda} = 0$. The system to solve is:
\begin{equation}
\begin{cases}
2\rho_{11} - \lambda = 0;\\
2\rho_{22} - \lambda = 0;\\
\cdots\\
2\rho_{NN} - \lambda = 0;\\
\rho_{11}+\rho_{22}+\dots+\rho_{NN} - 1 = 0.
\end{cases}
\end{equation}
The solution of this system is $\rho_{11} =\rho_{11} = \dots = \rho_{NN} = 1/N$.

\section{Poincar\'e map of the Classical Kicked Top}
\label{app:Derivation CKT map}
To obtain the classical limit of the QKT we introduce the normalised vector $\boldsymbol{X} = \Exp{\boldsymbol{J}}/j$ and take $j\rightarrow\infty$ (this is equivalent to taking the thermodynamic limit $N\rightarrow \infty$). Substituting this variable in to Eq.~(\ref{eq:heisenbergTop}), we obtain the classical map
\begin{equation}
\begin{split}
	X_{n+1} &= X_n\cos[\beta(Y_n\sin\alpha + Z_n\cos\alpha)]\\
		 &- (Y_n\cos\alpha - Z_n\sin\alpha)\sin[\beta(Y_n\sin\alpha + Z_n\cos\alpha)]~, \\
	Y_{n+1} &= X_n\sin[\beta(Y_n\sin\alpha + Z_n\cos\alpha)]\\
		&+ (Y_n\cos\alpha - Z_n\sin\alpha)\cos[\beta(Y_n\sin\alpha + Z_n\cos\alpha)]~, \\
	Z_{n+1} &= Y_n\sin\alpha + Z_n\cos\alpha~.
\end{split}
\label{eq:cTop}
\end{equation}
The normalised angular momentum vector can be parameterised in polar coordinates, $\boldsymbol{X} = (\sin\Theta \cos\Phi,\sin\Theta \sin\Phi,\cos\Theta)$, to give a two-dimensional classical phase space, in the form of a Poincar\'e map.

\section{Derivation of the tangent space of the CKT}
\label{app:Derivation of tangent space of CKT}

The procedure to obtain the KSE is as follows. The generalised iterative map $\boldsymbol{x}_{n+1} = f(\boldsymbol{x}_n)$ of Eq.~(\ref{eq:cTop}), is linearised to give its associated tangent map $\delta \boldsymbol{x}_{n+1} = f(\boldsymbol{x}_n + \delta \boldsymbol{x}_n) - f(\boldsymbol{x}_n)$, where $\boldsymbol{x}_n = (X_n,Y_n,Z_n)$. The tangent map is then rescaled, $\delta \boldsymbol{x}_{n} \rightarrow \delta \boldsymbol{x}_{n}/l_n$,  before being fed back at each iteration. 
 
We explicitly do the calculation in detail for the coordinate $X$ ($Y$ and $Z$ follow in a similar fashion). 

%\begin{widetext}	
	\[
	\begin{split}
	dX_{n+1} &= f(X_n + \delta X_n) - f(X_n) \\
			&= (X_n + dX_n)\cos\big\{\beta\big[(Y_n+dY_n)\sin\alpha + (Z_n+dZ_n)\cos\alpha\big]\big\}\\
		    &\quad- \big[(Y_n+dY_n)\cos\alpha - (Z_n+dZ_n)\sin\alpha\big]\sin\big\{\beta\big[(Y_n+dY_n)\sin\alpha + (Z_n + dZ_n)\cos\alpha\big]\big\}\\
		    &\quad - X_n\cos[\beta(Y_n\sin\alpha + Z_n\cos\alpha)] + (Y_n\cos\alpha - Z_n\sin\alpha)\sin[\beta(Y_n\sin\alpha + Z_n\cos\alpha)].
	\end{split}
	\]
	%\end{widetext}
	Expanding the expression and keeping only the terms up to the first order in $dX_n$, $dY_n$ and $dZ_n$, we obtain,	
	\[
	\begin{split}
		dX_{n+1} &= X_n \cos\big[(\beta Y_n  \sin\alpha+ \beta Z_n \cos\alpha ) +  (\beta  dY_n \sin\alpha + \beta dZ_n \cos\alpha)\big] + dX_n\cos\big[\beta  Y_n \sin\alpha+ \beta  Z_n \cos\alpha \big]\\
		&- Y_n\cos\alpha\sin\big[(\beta Y_n  \sin\alpha+ \beta  Z_n \cos\alpha) + (\beta  dY_n \sin\alpha+ \beta  dZ_n \cos\alpha)\big]- dY_n\cos\alpha\sin\big[\beta  Y_n \sin\alpha+ \beta Z_n \cos\alpha \big]\\
	& + Z_n \sin\alpha\sin\big[(\beta  Y_n  \sin\alpha+ \beta Z_n \cos\alpha ) +  (\beta  dY_n  \sin\alpha+ \beta  dZ_n \cos\alpha )\big]+ dZ_n\sin\alpha\sin\big[\beta  Y_n  \sin\alpha+ \beta  Z_n \cos\alpha\big]\\
	&  - X_n\cos\big[\beta  Y_n \sin\alpha+ \beta  Z_n \cos\alpha \big] + Y_n\cos\alpha\sin\big[\beta  Y_n \sin\alpha + \beta  Z_n \cos\alpha \big]\\
	& - Z_n\sin\alpha\sin\big[\beta  Y_n \sin\alpha + \beta  Z_n \cos\alpha \big].
	\end{split}
	\]	
Using the relation $\cos(x+dx) - \cos(dx) = -\sin(x)dx$ we get
		
	\[
	\begin{split}
	    dX_{n+1} &= - X_n \sin\big(\beta  Y_n \sin\alpha+ \beta Z_n \cos\alpha \big)d(\beta  Y_n \sin\alpha+ \beta  Z_n\cos\alpha) \\
	& - Y_n\cos\alpha\cos\big(\beta  Y_n \sin\alpha+ \beta  Z_n \cos\alpha \big)d(\beta  Y_n \sin\alpha + \beta Z_n \cos\alpha)\\
	& + Z_n\sin\alpha\cos\big(\beta  Y_n \sin\alpha + \beta  Z_n \cos\alpha \big)d(\beta  Y_n \sin\alpha + \beta  Z_n \cos\alpha )\\
	& + dX_n\cos\big[\beta  Y_n \sin\alpha + \beta  Z_n \cos\alpha \big]- dY_n\cos\alpha\sin\big[\beta  Y_n  \sin\alpha + \beta Z_n \cos\alpha \big] + dZ_n\sin\alpha\sin\big[\beta  Y_n \sin\alpha + \beta  Z_n \cos\alpha \big].
	\end{split}
	\]
%\end{widetext}	
Finally, grouping the $dX_n$, $dY_n$ and $dZ_n$ terms, we write down the tangent map in its standard linear form $dx_{n+1} = L(x_n)dx_n$:
\begin{equation} 
\begin{split}
	 \delta X_{n+1}&= \delta X_n\cos\gamma_n  \\
	 +& \delta Y_n\big[-X_n\beta\sin\alpha\sin \gamma_n   - Y_n\beta\sin\alpha\cos\alpha\cos \gamma_n  \\ 
	 +& Z_n\beta\sin^2\alpha\cos \gamma_n   - \cos\alpha\sin \gamma_n  \big]\\
	 +& \delta Z_n\big[-X_n\beta\cos\alpha\sin \gamma_n   - Y_n\beta\cos^2\alpha\cos \gamma_n  \\ 
	 +& Z_n\beta\cos\alpha\sin\alpha\cos \gamma_n   + \sin\alpha\sin \gamma_n  \big],\\
	 \delta Y_{n+1} &= \delta X_n\sin \gamma_n  \\
	 +& \delta Y_n\big[X_n\beta\sin\alpha\cos \gamma_n   - Y_n\beta\sin\alpha\cos\alpha\sin \gamma_n  \\ 
	 +& Z_n\beta\sin^2\alpha\sin \gamma_n   + \cos\alpha\cos \gamma_n  \big]\\
	 +& \delta Z_n\big[X_n\beta\cos\alpha\cos \gamma_n   - Y_n\beta\cos^2\alpha\sin \gamma_n  \\ 
	 +& Z_n\beta\cos\alpha\sin\alpha\sin \gamma_n   + \cos\alpha\cos \gamma_n  \big],\\
	 \delta Z_{n+1}&= \delta Y_n\sin\alpha + \delta Z_n \cos\alpha,
\end{split}
\label{eq:top_tangent}
\end{equation}
where $\gamma_n \equiv  Y_n \beta\sin\alpha + Z_n \beta\cos\alpha$. 

% As far as we are aware, this is the first time tangent space of the CKT has been written down, and the full phase space KSE calculated.

\section{Derivation of the tangent space of the CKR}
\label{app:Derivation of tangent space of CKR}

The generalised iterative map $\boldsymbol{x}_{n+1} = f(\boldsymbol{x}_n)$ of Eq.~(\ref{eq:cRotor}), is linearised to give its associated tangent map
$\delta \boldsymbol{x}_{n+1} = f(\boldsymbol{x}_n + \delta \boldsymbol{x}_n) - f(\boldsymbol{x}_n)$, where $\boldsymbol{x}_n = (\Phi_n,P_n)$.
\begin{equation}
	\begin{split}
	\delta P_{n+1} &= P_n + \delta P_n + K\sin(\Phi_n + \delta \Phi_n) - \big(P_n + K\sin\Phi_n\big)\\
	& =  \delta P_n + K\cos(\Phi_n)\delta \Phi_n .
	\end{split}
\end{equation}
\begin{equation}
	\begin{split}
	\delta \Phi_{n+1} &= \Phi_n + \delta\Phi_n + \frac{P_{n+1} + \delta P_{n+1}}{I} - \Phi_n - \frac{P_{n+1}}{I}\\
	& = \delta\Phi_n + \frac{\delta P_{n+1}}{I} = \delta\Phi_n + \frac{\delta P_n + K\cos(\Phi_n)\delta\Phi_n}{I} \\
	& = \big(1+ \frac{K}{I}\cos\Phi_n\big) \delta\Phi_n+ \frac{\delta P_n}{I}.
	\end{split}
\end{equation}
\end{widetext}

\bibliography{chaos2}

%merlin.mbs apsrev4-1.bst 2010-07-25 4.21a (PWD, AO, DPC) hacked
%Control: key (0)
%Control: author (0) dotless jnrlst
%Control: editor formatted (1) identically to author
%Control: production of article title (0) allowed
%Control: page (1) range
%Control: year (0) verbatim
%Control: production of eprint (0) enabled
\begin{thebibliography}{49}%
\makeatletter
\providecommand \@ifxundefined [1]{%
 \@ifx{#1\undefined}
}%
\providecommand \@ifnum [1]{%
 \ifnum #1\expandafter \@firstoftwo
 \else \expandafter \@secondoftwo
 \fi
}%
\providecommand \@ifx [1]{%
 \ifx #1\expandafter \@firstoftwo
 \else \expandafter \@secondoftwo
 \fi
}%
\providecommand \natexlab [1]{#1}%
\providecommand \enquote  [1]{``#1''}%
\providecommand \bibnamefont  [1]{#1}%
\providecommand \bibfnamefont [1]{#1}%
\providecommand \citenamefont [1]{#1}%
\providecommand \href@noop [0]{\@secondoftwo}%
\providecommand \href [0]{\begingroup \@sanitize@url \@href}%
\providecommand \@href[1]{\@@startlink{#1}\@@href}%
\providecommand \@@href[1]{\endgroup#1\@@endlink}%
\providecommand \@sanitize@url [0]{\catcode `\\12\catcode `\$12\catcode
  `\&12\catcode `\#12\catcode `\^12\catcode `\_12\catcode `\%12\relax}%
\providecommand \@@startlink[1]{}%
\providecommand \@@endlink[0]{}%
\providecommand \url  [0]{\begingroup\@sanitize@url \@url }%
\providecommand \@url [1]{\endgroup\@href {#1}{\urlprefix }}%
\providecommand \urlprefix  [0]{URL }%
\providecommand \Eprint [0]{\href }%
\providecommand \doibase [0]{http://dx.doi.org/}%
\providecommand \selectlanguage [0]{\@gobble}%
\providecommand \bibinfo  [0]{\@secondoftwo}%
\providecommand \bibfield  [0]{\@secondoftwo}%
\providecommand \translation [1]{[#1]}%
\providecommand \BibitemOpen [0]{}%
\providecommand \bibitemStop [0]{}%
\providecommand \bibitemNoStop [0]{.\EOS\space}%
\providecommand \EOS [0]{\spacefactor3000\relax}%
\providecommand \BibitemShut  [1]{\csname bibitem#1\endcsname}%
\let\auto@bib@innerbib\@empty
%</preamble>
\bibitem [{\citenamefont {Arnol'd}(2013)}]{arnold78}%
  \BibitemOpen
  \bibfield  {author} {\bibinfo {author} {\bibfnamefont {V.~I}\ \bibnamefont
  {Arnol'd}},\ }\href@noop {} {\emph {\bibinfo {title} {Mathematical methods of
  classical mechanics}}},\ Vol.~\bibinfo {volume} {60}\ (\bibinfo  {publisher}
  {Springer Science \& Business Media},\ \bibinfo {year} {2013})\BibitemShut
  {NoStop}%
\bibitem [{\citenamefont {Reichl}(2013)}]{reichl13}%
  \BibitemOpen
  \bibfield  {author} {\bibinfo {author} {\bibfnamefont {L.}~\bibnamefont
  {Reichl}},\ }\href@noop {} {\emph {\bibinfo {title} {The transition to chaos:
  conservative classical systems and quantum manifestations}}}\ (\bibinfo
  {publisher} {Springer Science \& Business Media},\ \bibinfo {year}
  {2013})\BibitemShut {NoStop}%
\bibitem [{\citenamefont {Haake}(2013)}]{haake13}%
  \BibitemOpen
  \bibfield  {author} {\bibinfo {author} {\bibfnamefont {F.}~\bibnamefont
  {Haake}},\ }\href@noop {} {\emph {\bibinfo {title} {Quantum signatures of
  chaos}}},\ Vol.~\bibinfo {volume} {54}\ (\bibinfo  {publisher} {Springer
  Science \& Business Media},\ \bibinfo {year} {2013})\BibitemShut {NoStop}%
\bibitem [{\citenamefont {Gutzwiller}(1990)}]{gutzwiller90}%
  \BibitemOpen
  \bibfield  {author} {\bibinfo {author} {\bibfnamefont {M.~C.}\ \bibnamefont
  {Gutzwiller}},\ }\href@noop {} {\emph {\bibinfo {title} {Chaos in classical
  and quantum mechanics, volume 1 of interdisciplinary applied mathematics}}}\
  (\bibinfo  {publisher} {Springer-Verlag, New York},\ \bibinfo {year}
  {1990})\BibitemShut {NoStop}%
\bibitem [{\citenamefont {Tabor}(1989)}]{tabor89}%
  \BibitemOpen
  \bibfield  {author} {\bibinfo {author} {\bibfnamefont {M.}~\bibnamefont
  {Tabor}},\ }\href@noop {} {\emph {\bibinfo {title} {Chaos and integrability
  in nonlinear dynamics: an introduction}}}\ (\bibinfo  {publisher} {Wiley},\
  \bibinfo {year} {1989})\BibitemShut {NoStop}%
\bibitem [{\citenamefont {Zyczkowski}(1990)}]{zyczkowski1990}%
  \BibitemOpen
  \bibfield  {author} {\bibinfo {author} {\bibfnamefont {K.}~\bibnamefont
  {Zyczkowski}},\ }\bibfield  {title} {\enquote {\bibinfo {title} {Indicators
  of quantum chaos based on eigenvector statistics},}\ }\href@noop {}
  {\bibfield  {journal} {\bibinfo  {journal} {Journal of Physics A:
  Mathematical and General}\ }\textbf {\bibinfo {volume} {23}},\ \bibinfo
  {pages} {4427} (\bibinfo {year} {1990})}\BibitemShut {NoStop}%
\bibitem [{\citenamefont {Ford}\ \emph {et~al.}(1991)\citenamefont {Ford},
  \citenamefont {Mantica},\ and\ \citenamefont {Ristow}}]{ford91}%
  \BibitemOpen
  \bibfield  {author} {\bibinfo {author} {\bibfnamefont {J.}~\bibnamefont
  {Ford}}, \bibinfo {author} {\bibfnamefont {G.}~\bibnamefont {Mantica}}, \
  and\ \bibinfo {author} {\bibfnamefont {G.~H.}\ \bibnamefont {Ristow}},\
  }\bibfield  {title} {\enquote {\bibinfo {title} {The arnol'd cat: Failure of
  the correspondence principle},}\ }\href {\doibase
  https://doi.org/10.1016/0167-2789(91)90012-X} {\bibfield  {journal} {\bibinfo
   {journal} {Physica D: Nonlinear Phenomena}\ }\textbf {\bibinfo {volume}
  {50}},\ \bibinfo {pages} {493 -- 520} (\bibinfo {year} {1991})}\BibitemShut
  {NoStop}%
\bibitem [{\citenamefont {Schack}\ and\ \citenamefont
  {Caves}(1996{\natexlab{a}})}]{schack96}%
  \BibitemOpen
  \bibfield  {author} {\bibinfo {author} {\bibfnamefont {R.}~\bibnamefont
  {Schack}}\ and\ \bibinfo {author} {\bibfnamefont {C.~M.}\ \bibnamefont
  {Caves}},\ }\bibfield  {title} {\enquote {\bibinfo {title} {Chaos for
  liouville probability densities},}\ }\href {\doibase
  10.1103/PhysRevE.53.3387} {\bibfield  {journal} {\bibinfo  {journal} {Phys.
  Rev. E}\ }\textbf {\bibinfo {volume} {53}},\ \bibinfo {pages} {3387--3401}
  (\bibinfo {year} {1996}{\natexlab{a}})}\BibitemShut {NoStop}%
\bibitem [{\citenamefont {Slomczy{\'n}ski}\ and\ \citenamefont
  {{\.Z}yczkowski}(1994)}]{slomczynski96}%
  \BibitemOpen
  \bibfield  {author} {\bibinfo {author} {\bibfnamefont {W.}~\bibnamefont
  {Slomczy{\'n}ski}}\ and\ \bibinfo {author} {\bibfnamefont {K.}~\bibnamefont
  {{\.Z}yczkowski}},\ }\bibfield  {title} {\enquote {\bibinfo {title} {Quantum
  chaos: an entropy approach},}\ }\href@noop {} {\bibfield  {journal} {\bibinfo
   {journal} {Journal of Mathematical Physics}\ }\textbf {\bibinfo {volume}
  {35}},\ \bibinfo {pages} {5674--5700} (\bibinfo {year} {1994})}\BibitemShut
  {NoStop}%
\bibitem [{\citenamefont {Schack}\ and\ \citenamefont
  {Caves}(1996{\natexlab{b}})}]{schack96a}%
  \BibitemOpen
  \bibfield  {author} {\bibinfo {author} {\bibfnamefont {R.}~\bibnamefont
  {Schack}}\ and\ \bibinfo {author} {\bibfnamefont {C.~M.}\ \bibnamefont
  {Caves}},\ }\bibfield  {title} {\enquote {\bibinfo {title}
  {Information-theoretic characterization of quantum chaos},}\ }\href {\doibase
  10.1103/PhysRevE.53.3257} {\bibfield  {journal} {\bibinfo  {journal} {Phys.
  Rev. E}\ }\textbf {\bibinfo {volume} {53}},\ \bibinfo {pages} {3257--3270}
  (\bibinfo {year} {1996}{\natexlab{b}})}\BibitemShut {NoStop}%
\bibitem [{\citenamefont {Zurek}\ and\ \citenamefont {Paz}(1994)}]{zurek94}%
  \BibitemOpen
  \bibfield  {author} {\bibinfo {author} {\bibfnamefont {W.~H.}\ \bibnamefont
  {Zurek}}\ and\ \bibinfo {author} {\bibfnamefont {J.~P.}\ \bibnamefont
  {Paz}},\ }\bibfield  {title} {\enquote {\bibinfo {title} {Decoherence, chaos,
  and the second law},}\ }\href {\doibase 10.1103/PhysRevLett.72.2508}
  {\bibfield  {journal} {\bibinfo  {journal} {Phys. Rev. Lett.}\ }\textbf
  {\bibinfo {volume} {72}},\ \bibinfo {pages} {2508--2511} (\bibinfo {year}
  {1994})}\BibitemShut {NoStop}%
\bibitem [{\citenamefont {Zurek}\ and\ \citenamefont {Paz}(1995)}]{zurek95}%
  \BibitemOpen
  \bibfield  {author} {\bibinfo {author} {\bibfnamefont {W.~H.}\ \bibnamefont
  {Zurek}}\ and\ \bibinfo {author} {\bibfnamefont {J.~P.}\ \bibnamefont
  {Paz}},\ }\bibfield  {title} {\enquote {\bibinfo {title} {Quantum chaos: a
  decoherent definition},}\ }\href {\doibase
  https://doi.org/10.1016/0167-2789(94)00271-Q} {\bibfield  {journal} {\bibinfo
   {journal} {Physica D: Nonlinear Phenomena}\ }\textbf {\bibinfo {volume}
  {83}},\ \bibinfo {pages} {300 -- 308} (\bibinfo {year} {1995})},\ \bibinfo
  {note} {quantum Complexity in Mesoscopic Systems}\BibitemShut {NoStop}%
\bibitem [{\citenamefont {Demkowicz-Dobrza\ifmmode~\acute{n}\else
  \'{n}\fi{}ski}\ and\ \citenamefont {Ku{\'{s}}}(2004)}]{Demkowicz2004}%
  \BibitemOpen
  \bibfield  {author} {\bibinfo {author} {\bibfnamefont {R.}~\bibnamefont
  {Demkowicz-Dobrza\ifmmode~\acute{n}\else \'{n}\fi{}ski}}\ and\ \bibinfo
  {author} {\bibfnamefont {M.}~\bibnamefont {Ku{\'{s}}}},\ }\bibfield  {title}
  {\enquote {\bibinfo {title} {Global entangling properties of the coupled
  kicked tops},}\ }\href {\doibase 10.1103/PhysRevE.70.066216} {\bibfield
  {journal} {\bibinfo  {journal} {Phys. Rev. E}\ }\textbf {\bibinfo {volume}
  {70}},\ \bibinfo {pages} {066216} (\bibinfo {year} {2004})}\BibitemShut
  {NoStop}%
\bibitem [{\citenamefont {Evans}(2004)}]{evans2004towards}%
  \BibitemOpen
  \bibfield  {author} {\bibinfo {author} {\bibfnamefont {L.~C.}\ \bibnamefont
  {Evans}},\ }\bibfield  {title} {\enquote {\bibinfo {title} {Towards a quantum
  analog of weak kam theory},}\ }\href@noop {} {\bibfield  {journal} {\bibinfo
  {journal} {Communications in Mathematical Physics}\ }\textbf {\bibinfo
  {volume} {244}},\ \bibinfo {pages} {311--334} (\bibinfo {year}
  {2004})}\BibitemShut {NoStop}%
\bibitem [{\citenamefont {Hose}\ \emph {et~al.}(1984)\citenamefont {Hose},
  \citenamefont {Taylor},\ and\ \citenamefont {Tip}}]{hose84}%
  \BibitemOpen
  \bibfield  {author} {\bibinfo {author} {\bibfnamefont {G.}~\bibnamefont
  {Hose}}, \bibinfo {author} {\bibfnamefont {H.~S.}\ \bibnamefont {Taylor}}, \
  and\ \bibinfo {author} {\bibfnamefont {A.}~\bibnamefont {Tip}},\ }\bibfield
  {title} {\enquote {\bibinfo {title} {A quantum kam-like theorem. ii.
  fundamentals of localisation in quantum theory for resonance states},}\
  }\href {http://stacks.iop.org/0305-4470/17/i=6/a=016} {\bibfield  {journal}
  {\bibinfo  {journal} {Journal of Physics A: Mathematical and General}\
  }\textbf {\bibinfo {volume} {17}},\ \bibinfo {pages} {1203} (\bibinfo {year}
  {1984})}\BibitemShut {NoStop}%
\bibitem [{\citenamefont {Brandino}\ \emph {et~al.}(2015)\citenamefont
  {Brandino}, \citenamefont {Caux},\ and\ \citenamefont {Konik}}]{brandino15}%
  \BibitemOpen
  \bibfield  {author} {\bibinfo {author} {\bibfnamefont {G.~P.}\ \bibnamefont
  {Brandino}}, \bibinfo {author} {\bibfnamefont {J.-S.}\ \bibnamefont {Caux}},
  \ and\ \bibinfo {author} {\bibfnamefont {R.~M.}\ \bibnamefont {Konik}},\
  }\bibfield  {title} {\enquote {\bibinfo {title} {Glimmers of a quantum kam
  theorem: insights from quantum quenches in one-dimensional bose gases},}\
  }\href {\doibase 10.1103/PhysRevX.5.041043} {\bibfield  {journal} {\bibinfo
  {journal} {Phys. Rev. X}\ }\textbf {\bibinfo {volume} {5}},\ \bibinfo {pages}
  {041043} (\bibinfo {year} {2015})}\BibitemShut {NoStop}%
\bibitem [{\citenamefont {Geisel}\ \emph {et~al.}(1986)\citenamefont {Geisel},
  \citenamefont {Radons},\ and\ \citenamefont {Rubner}}]{geisel86}%
  \BibitemOpen
  \bibfield  {author} {\bibinfo {author} {\bibfnamefont {T.}~\bibnamefont
  {Geisel}}, \bibinfo {author} {\bibfnamefont {G.}~\bibnamefont {Radons}}, \
  and\ \bibinfo {author} {\bibfnamefont {J.}~\bibnamefont {Rubner}},\
  }\bibfield  {title} {\enquote {\bibinfo {title} {Kolmogorov-arnol'd-moser
  barriers in the quantum dynamics of chaotic systems},}\ }\href {\doibase
  10.1103/PhysRevLett.57.2883} {\bibfield  {journal} {\bibinfo  {journal}
  {Phys. Rev. Lett.}\ }\textbf {\bibinfo {volume} {57}},\ \bibinfo {pages}
  {2883--2886} (\bibinfo {year} {1986})}\BibitemShut {NoStop}%
\bibitem [{\citenamefont {Pesin}(1977)}]{pesin77}%
  \BibitemOpen
  \bibfield  {author} {\bibinfo {author} {\bibfnamefont {Y.~B.}\ \bibnamefont
  {Pesin}},\ }\bibfield  {title} {\enquote {\bibinfo {title} {Characteristic
  lyapunov exponents and smooth ergodic theory},}\ }\href
  {http://stacks.iop.org/0036-0279/32/i=4/a=R02} {\bibfield  {journal}
  {\bibinfo  {journal} {Russian Mathematical Surveys}\ }\textbf {\bibinfo
  {volume} {32}},\ \bibinfo {pages} {55} (\bibinfo {year} {1977})}\BibitemShut
  {NoStop}%
\bibitem [{\citenamefont {Miller}\ and\ \citenamefont
  {Sarkar}(1998)}]{miller98}%
  \BibitemOpen
  \bibfield  {author} {\bibinfo {author} {\bibfnamefont {P.~A.}\ \bibnamefont
  {Miller}}\ and\ \bibinfo {author} {\bibfnamefont {S.}~\bibnamefont
  {Sarkar}},\ }\bibfield  {title} {\enquote {\bibinfo {title} {Fingerprints of
  classical instability in open quantum dynamics},}\ }\href@noop {} {\bibfield
  {journal} {\bibinfo  {journal} {Physical Review E}\ }\textbf {\bibinfo
  {volume} {58}},\ \bibinfo {pages} {4217} (\bibinfo {year}
  {1998})}\BibitemShut {NoStop}%
\bibitem [{\citenamefont {Zarum}\ and\ \citenamefont {Sarkar}(1998)}]{zarum98}%
  \BibitemOpen
  \bibfield  {author} {\bibinfo {author} {\bibfnamefont {R.}~\bibnamefont
  {Zarum}}\ and\ \bibinfo {author} {\bibfnamefont {S.}~\bibnamefont {Sarkar}},\
  }\bibfield  {title} {\enquote {\bibinfo {title} {Quantum-classical
  correspondence of entropy contours in the transition to chaos},}\ }\href
  {\doibase 10.1103/PhysRevE.57.5467} {\bibfield  {journal} {\bibinfo
  {journal} {Phys. Rev. E}\ }\textbf {\bibinfo {volume} {57}},\ \bibinfo
  {pages} {5467--5471} (\bibinfo {year} {1998})}\BibitemShut {NoStop}%
\bibitem [{\citenamefont {Furuya}\ \emph {et~al.}(1998)\citenamefont {Furuya},
  \citenamefont {Nemes},\ and\ \citenamefont {Pellegrino}}]{furuya98}%
  \BibitemOpen
  \bibfield  {author} {\bibinfo {author} {\bibfnamefont {K.}~\bibnamefont
  {Furuya}}, \bibinfo {author} {\bibfnamefont {M.~C.}\ \bibnamefont {Nemes}}, \
  and\ \bibinfo {author} {\bibfnamefont {G.~Q.}\ \bibnamefont {Pellegrino}},\
  }\bibfield  {title} {\enquote {\bibinfo {title} {Quantum dynamical
  manifestation of chaotic behavior in the process of entanglement},}\ }\href
  {\doibase 10.1103/PhysRevLett.80.5524} {\bibfield  {journal} {\bibinfo
  {journal} {Phys. Rev. Lett.}\ }\textbf {\bibinfo {volume} {80}},\ \bibinfo
  {pages} {5524--5527} (\bibinfo {year} {1998})}\BibitemShut {NoStop}%
\bibitem [{\citenamefont {Wang}\ \emph {et~al.}(2004)\citenamefont {Wang},
  \citenamefont {Ghose}, \citenamefont {Sanders},\ and\ \citenamefont
  {Hu}}]{wang04}%
  \BibitemOpen
  \bibfield  {author} {\bibinfo {author} {\bibfnamefont {X.}~\bibnamefont
  {Wang}}, \bibinfo {author} {\bibfnamefont {S.}~\bibnamefont {Ghose}},
  \bibinfo {author} {\bibfnamefont {B.~C.}\ \bibnamefont {Sanders}}, \ and\
  \bibinfo {author} {\bibfnamefont {B.}~\bibnamefont {Hu}},\ }\bibfield
  {title} {\enquote {\bibinfo {title} {Entanglement as a signature of quantum
  chaos},}\ }\href {\doibase 10.1103/PhysRevE.70.016217} {\bibfield  {journal}
  {\bibinfo  {journal} {Phys. Rev. E}\ }\textbf {\bibinfo {volume} {70}},\
  \bibinfo {pages} {016217} (\bibinfo {year} {2004})}\BibitemShut {NoStop}%
\bibitem [{\citenamefont {Ghose}\ \emph {et~al.}(2008)\citenamefont {Ghose},
  \citenamefont {Stock}, \citenamefont {Jessen}, \citenamefont {Lal},\ and\
  \citenamefont {Silberfarb}}]{ghose08}%
  \BibitemOpen
  \bibfield  {author} {\bibinfo {author} {\bibfnamefont {S.}~\bibnamefont
  {Ghose}}, \bibinfo {author} {\bibfnamefont {R.}~\bibnamefont {Stock}},
  \bibinfo {author} {\bibfnamefont {P.}~\bibnamefont {Jessen}}, \bibinfo
  {author} {\bibfnamefont {R.}~\bibnamefont {Lal}}, \ and\ \bibinfo {author}
  {\bibfnamefont {A.}~\bibnamefont {Silberfarb}},\ }\bibfield  {title}
  {\enquote {\bibinfo {title} {Chaos, entanglement, and decoherence in the
  quantum kicked top},}\ }\href {\doibase 10.1103/PhysRevA.78.042318}
  {\bibfield  {journal} {\bibinfo  {journal} {Phys. Rev. A}\ }\textbf {\bibinfo
  {volume} {78}},\ \bibinfo {pages} {042318} (\bibinfo {year}
  {2008})}\BibitemShut {NoStop}%
\bibitem [{\citenamefont {Ruebeck}\ \emph {et~al.}(2017)\citenamefont
  {Ruebeck}, \citenamefont {Lin},\ and\ \citenamefont
  {Pattanayak}}]{ruebeck2017}%
  \BibitemOpen
  \bibfield  {author} {\bibinfo {author} {\bibfnamefont {J.~B.}\ \bibnamefont
  {Ruebeck}}, \bibinfo {author} {\bibfnamefont {J.}~\bibnamefont {Lin}}, \ and\
  \bibinfo {author} {\bibfnamefont {A.~K.}\ \bibnamefont {Pattanayak}},\
  }\bibfield  {title} {\enquote {\bibinfo {title} {Entanglement and its
  relationship to classical dynamics},}\ }\href@noop {} {\bibfield  {journal}
  {\bibinfo  {journal} {Physical Review E}\ }\textbf {\bibinfo {volume} {95}},\
  \bibinfo {pages} {062222} (\bibinfo {year} {2017})}\BibitemShut {NoStop}%
\bibitem [{\citenamefont {Kumari}\ and\ \citenamefont
  {Ghose}(2018{\natexlab{a}})}]{kumari2018orbits}%
  \BibitemOpen
  \bibfield  {author} {\bibinfo {author} {\bibfnamefont {M.}~\bibnamefont
  {Kumari}}\ and\ \bibinfo {author} {\bibfnamefont {S.}~\bibnamefont {Ghose}},\
  }\bibfield  {title} {\enquote {\bibinfo {title} {Quantum-classical
  correspondence in the vicinity of periodic orbits},}\ }\href@noop {}
  {\bibfield  {journal} {\bibinfo  {journal} {Physical Review E}\ }\textbf
  {\bibinfo {volume} {97}},\ \bibinfo {pages} {052209} (\bibinfo {year}
  {2018}{\natexlab{a}})}\BibitemShut {NoStop}%
\bibitem [{\citenamefont {Kumari}\ and\ \citenamefont
  {Ghose}(2018{\natexlab{b}})}]{kumari2018untangling}%
  \BibitemOpen
  \bibfield  {author} {\bibinfo {author} {\bibfnamefont {M.}~\bibnamefont
  {Kumari}}\ and\ \bibinfo {author} {\bibfnamefont {S.}~\bibnamefont {Ghose}},\
  }\bibfield  {title} {\enquote {\bibinfo {title} {Untangling entanglement and
  chaos},}\ }\href@noop {} {\bibfield  {journal} {\bibinfo  {journal} {arXiv
  preprint arXiv:1806.10545}\ } (\bibinfo {year}
  {2018}{\natexlab{b}})}\BibitemShut {NoStop}%
\bibitem [{\citenamefont {Neill}\ \emph {et~al.}(2016)\citenamefont {Neill},
  \citenamefont {Roushan}, \citenamefont {Fang}, \citenamefont {Chen},
  \citenamefont {Kolodrubetz}, \citenamefont {Chen}, \citenamefont {Megrant},
  \citenamefont {Barends}, \citenamefont {Campbell}, \citenamefont {Chiaro}
  \emph {et~al.}}]{neill16}%
  \BibitemOpen
  \bibfield  {author} {\bibinfo {author} {\bibfnamefont {C.}~\bibnamefont
  {Neill}}, \bibinfo {author} {\bibfnamefont {P.}~\bibnamefont {Roushan}},
  \bibinfo {author} {\bibfnamefont {M.}~\bibnamefont {Fang}}, \bibinfo {author}
  {\bibfnamefont {Y.}~\bibnamefont {Chen}}, \bibinfo {author} {\bibfnamefont
  {M.}~\bibnamefont {Kolodrubetz}}, \bibinfo {author} {\bibfnamefont
  {Z.}~\bibnamefont {Chen}}, \bibinfo {author} {\bibfnamefont {A.}~\bibnamefont
  {Megrant}}, \bibinfo {author} {\bibfnamefont {R.}~\bibnamefont {Barends}},
  \bibinfo {author} {\bibfnamefont {B.}~\bibnamefont {Campbell}}, \bibinfo
  {author} {\bibfnamefont {B.}~\bibnamefont {Chiaro}},  \emph {et~al.},\
  }\bibfield  {title} {\enquote {\bibinfo {title} {Ergodic dynamics and
  thermalization in an isolated quantum system},}\ }\href@noop {} {\bibfield
  {journal} {\bibinfo  {journal} {Nature Physics}\ } (\bibinfo {year}
  {2016})}\BibitemShut {NoStop}%
\bibitem [{\citenamefont {Lombardi}\ and\ \citenamefont
  {Matzkin}(2011)}]{lombardi11}%
  \BibitemOpen
  \bibfield  {author} {\bibinfo {author} {\bibfnamefont {M.}~\bibnamefont
  {Lombardi}}\ and\ \bibinfo {author} {\bibfnamefont {A.}~\bibnamefont
  {Matzkin}},\ }\bibfield  {title} {\enquote {\bibinfo {title} {Entanglement
  and chaos in the kicked top},}\ }\href {\doibase 10.1103/PhysRevE.83.016207}
  {\bibfield  {journal} {\bibinfo  {journal} {Phys. Rev. E}\ }\textbf {\bibinfo
  {volume} {83}},\ \bibinfo {pages} {016207} (\bibinfo {year}
  {2011})}\BibitemShut {NoStop}%
\bibitem [{\citenamefont {Lombardi}\ and\ \citenamefont
  {Matzkin}(2015)}]{lombardi15a}%
  \BibitemOpen
  \bibfield  {author} {\bibinfo {author} {\bibfnamefont {M.}~\bibnamefont
  {Lombardi}}\ and\ \bibinfo {author} {\bibfnamefont {A.}~\bibnamefont
  {Matzkin}},\ }\bibfield  {title} {\enquote {\bibinfo {title} {Reply to
  ``comment on `entanglement and chaos in the kicked top' ''},}\ }\href
  {\doibase 10.1103/PhysRevE.92.036902} {\bibfield  {journal} {\bibinfo
  {journal} {Phys. Rev. E}\ }\textbf {\bibinfo {volume} {92}},\ \bibinfo
  {pages} {036902} (\bibinfo {year} {2015})}\BibitemShut {NoStop}%
\bibitem [{\citenamefont {Madhok}(2015)}]{madhok15}%
  \BibitemOpen
  \bibfield  {author} {\bibinfo {author} {\bibfnamefont {V.}~\bibnamefont
  {Madhok}},\ }\bibfield  {title} {\enquote {\bibinfo {title} {Comment on
  ``entanglement and chaos in the kicked top''},}\ }\href {\doibase
  10.1103/PhysRevE.92.036901} {\bibfield  {journal} {\bibinfo  {journal} {Phys.
  Rev. E}\ }\textbf {\bibinfo {volume} {92}},\ \bibinfo {pages} {036901}
  (\bibinfo {year} {2015})}\BibitemShut {NoStop}%
\bibitem [{\citenamefont {Haake}\ and\ \citenamefont
  {Shepelyansky}(1988)}]{haake88}%
  \BibitemOpen
  \bibfield  {author} {\bibinfo {author} {\bibfnamefont {F.}~\bibnamefont
  {Haake}}\ and\ \bibinfo {author} {\bibfnamefont {D.~L.}\ \bibnamefont
  {Shepelyansky}},\ }\bibfield  {title} {\enquote {\bibinfo {title} {The kicked
  rotator as a limit of the kicked top},}\ }\href
  {http://stacks.iop.org/0295-5075/5/i=8/a=001} {\bibfield  {journal} {\bibinfo
   {journal} {EPL (Europhysics Letters)}\ }\textbf {\bibinfo {volume} {5}},\
  \bibinfo {pages} {671} (\bibinfo {year} {1988})}\BibitemShut {NoStop}%
\bibitem [{\citenamefont {Haake}\ \emph {et~al.}(1987)\citenamefont {Haake},
  \citenamefont {Ku{\'{s}}},\ and\ \citenamefont {Scharf}}]{haake86}%
  \BibitemOpen
  \bibfield  {author} {\bibinfo {author} {\bibfnamefont {F}~\bibnamefont
  {Haake}}, \bibinfo {author} {\bibfnamefont {M.}~\bibnamefont {Ku{\'{s}}}}, \
  and\ \bibinfo {author} {\bibfnamefont {R.}~\bibnamefont {Scharf}},\
  }\bibfield  {title} {\enquote {\bibinfo {title} {Classical and quantum chaos
  for a kicked top},}\ }\href {\doibase 10.1007/BF01303727} {\bibfield
  {journal} {\bibinfo  {journal} {Zeitschrift f{\"u}r Physik B Condensed
  Matter}\ }\textbf {\bibinfo {volume} {65}},\ \bibinfo {pages} {381--395}
  (\bibinfo {year} {1987})}\BibitemShut {NoStop}%
\bibitem [{\citenamefont {Kolmogorov}(1958)}]{kolmogorov58}%
  \BibitemOpen
  \bibfield  {author} {\bibinfo {author} {\bibfnamefont {A.~N.}\ \bibnamefont
  {Kolmogorov}},\ }\bibfield  {title} {\enquote {\bibinfo {title} {A new metric
  invariant of transient dynamical systems and automorphisms in lebesgue
  spaces},}\ }in\ \href@noop {} {\emph {\bibinfo {booktitle} {Dokl. Akad. Nauk
  SSSR (NS)}}},\ Vol.\ \bibinfo {volume} {119}\ (\bibinfo {year} {1958})\
  p.~\bibinfo {pages} {2}\BibitemShut {NoStop}%
\bibitem [{\citenamefont {Kolmogorov}(1959)}]{kolmogorov59}%
  \BibitemOpen
  \bibfield  {author} {\bibinfo {author} {\bibfnamefont {A.~N.}\ \bibnamefont
  {Kolmogorov}},\ }\bibfield  {title} {\enquote {\bibinfo {title} {Entropy per
  unit time as a metric invariant of automorphisms},}\ }in\ \href@noop {}
  {\emph {\bibinfo {booktitle} {Dokl. Akad. Nauk SSSR}}},\ Vol.\ \bibinfo
  {volume} {124}\ (\bibinfo {year} {1959})\ pp.\ \bibinfo {pages}
  {754--755}\BibitemShut {NoStop}%
\bibitem [{\citenamefont {Lakshminarayan}(2001)}]{lakshminarayan01}%
  \BibitemOpen
  \bibfield  {author} {\bibinfo {author} {\bibfnamefont {A.}~\bibnamefont
  {Lakshminarayan}},\ }\bibfield  {title} {\enquote {\bibinfo {title}
  {Entangling power of quantized chaotic systems},}\ }\href {\doibase
  10.1103/PhysRevE.64.036207} {\bibfield  {journal} {\bibinfo  {journal} {Phys.
  Rev. E}\ }\textbf {\bibinfo {volume} {64}},\ \bibinfo {pages} {036207}
  (\bibinfo {year} {2001})}\BibitemShut {NoStop}%
\bibitem [{\citenamefont {Berry}\ and\ \citenamefont {Tabor}(1977)}]{berry77}%
  \BibitemOpen
  \bibfield  {author} {\bibinfo {author} {\bibfnamefont {M.~V.}\ \bibnamefont
  {Berry}}\ and\ \bibinfo {author} {\bibfnamefont {M.}~\bibnamefont {Tabor}},\
  }\bibfield  {title} {\enquote {\bibinfo {title} {Level clustering in the
  regular spectrum},}\ }\href {\doibase 10.1098/rspa.1977.0140} {\bibfield
  {journal} {\bibinfo  {journal} {Proceedings of the Royal Society of London A:
  Mathematical, Physical and Engineering Sciences}\ }\textbf {\bibinfo {volume}
  {356}},\ \bibinfo {pages} {375--394} (\bibinfo {year} {1977})}\BibitemShut
  {NoStop}%
\bibitem [{\citenamefont {Brandino}\ \emph {et~al.}(2010)\citenamefont
  {Brandino}, \citenamefont {Konik},\ and\ \citenamefont
  {Mussardo}}]{brandino10}%
  \BibitemOpen
  \bibfield  {author} {\bibinfo {author} {\bibfnamefont {G.~P.}\ \bibnamefont
  {Brandino}}, \bibinfo {author} {\bibfnamefont {R.~M.}\ \bibnamefont {Konik}},
  \ and\ \bibinfo {author} {\bibfnamefont {G.}~\bibnamefont {Mussardo}},\
  }\bibfield  {title} {\enquote {\bibinfo {title} {Energy level distribution of
  perturbed conformal field theories},}\ }\href
  {http://stacks.iop.org/1742-5468/2010/i=07/a=P07013} {\bibfield  {journal}
  {\bibinfo  {journal} {Journal of Statistical Mechanics: Theory and
  Experiment}\ }\textbf {\bibinfo {volume} {2010}},\ \bibinfo {pages} {P07013}
  (\bibinfo {year} {2010})}\BibitemShut {NoStop}%
\bibitem [{\citenamefont {Rigol}(2009{\natexlab{a}})}]{rigol09}%
  \BibitemOpen
  \bibfield  {author} {\bibinfo {author} {\bibfnamefont {M.}~\bibnamefont
  {Rigol}},\ }\bibfield  {title} {\enquote {\bibinfo {title} {Breakdown of
  thermalization in finite one-dimensional systems},}\ }\href {\doibase
  10.1103/PhysRevLett.103.100403} {\bibfield  {journal} {\bibinfo  {journal}
  {Phys. Rev. Lett.}\ }\textbf {\bibinfo {volume} {103}},\ \bibinfo {pages}
  {100403} (\bibinfo {year} {2009}{\natexlab{a}})}\BibitemShut {NoStop}%
\bibitem [{\citenamefont {Rigol}(2009{\natexlab{b}})}]{rigol09a}%
  \BibitemOpen
  \bibfield  {author} {\bibinfo {author} {\bibfnamefont {M.}~\bibnamefont
  {Rigol}},\ }\bibfield  {title} {\enquote {\bibinfo {title} {Quantum quenches
  and thermalization in one-dimensional fermionic systems},}\ }\href {\doibase
  10.1103/PhysRevA.80.053607} {\bibfield  {journal} {\bibinfo  {journal} {Phys.
  Rev. A}\ }\textbf {\bibinfo {volume} {80}},\ \bibinfo {pages} {053607}
  (\bibinfo {year} {2009}{\natexlab{b}})}\BibitemShut {NoStop}%
\bibitem [{\citenamefont {Percival}(1973)}]{percival73}%
  \BibitemOpen
  \bibfield  {author} {\bibinfo {author} {\bibfnamefont {I.~C.}\ \bibnamefont
  {Percival}},\ }\bibfield  {title} {\enquote {\bibinfo {title} {Regular and
  irregular spectra},}\ }\href {http://stacks.iop.org/0022-3700/6/i=9/a=002}
  {\bibfield  {journal} {\bibinfo  {journal} {Journal of Physics B: Atomic and
  Molecular Physics}\ }\textbf {\bibinfo {volume} {6}},\ \bibinfo {pages}
  {L229} (\bibinfo {year} {1973})}\BibitemShut {NoStop}%
\bibitem [{\citenamefont {Berry}(1977)}]{berry77a}%
  \BibitemOpen
  \bibfield  {author} {\bibinfo {author} {\bibfnamefont {M.~V.}\ \bibnamefont
  {Berry}},\ }\bibfield  {title} {\enquote {\bibinfo {title} {Regular and
  irregular semiclassical wavefunctions},}\ }\href
  {http://stacks.iop.org/0305-4470/10/i=12/a=016} {\bibfield  {journal}
  {\bibinfo  {journal} {Journal of Physics A: Mathematical and General}\
  }\textbf {\bibinfo {volume} {10}},\ \bibinfo {pages} {2083} (\bibinfo {year}
  {1977})}\BibitemShut {NoStop}%
\bibitem [{\citenamefont {Voros}(1979)}]{voros79}%
  \BibitemOpen
  \bibfield  {author} {\bibinfo {author} {\bibfnamefont {A.}~\bibnamefont
  {Voros}},\ }\bibfield  {title} {\enquote {\bibinfo {title} {Semi-classical
  ergodicity of quantum eigenstates in the wigner representation},}\ }in\
  \href@noop {} {\emph {\bibinfo {booktitle} {Stochastic Behavior in Classical
  and Quantum Hamiltonian Systems}}},\ \bibinfo {editor} {edited by\ \bibinfo
  {editor} {\bibfnamefont {G.}~\bibnamefont {Casati}}\ and\ \bibinfo {editor}
  {\bibfnamefont {J.}~\bibnamefont {Ford}}}\ (\bibinfo  {publisher} {Springer
  Berlin Heidelberg},\ \bibinfo {address} {Berlin, Heidelberg},\ \bibinfo
  {year} {1979})\ pp.\ \bibinfo {pages} {326--333}\BibitemShut {NoStop}%
\bibitem [{\citenamefont {Hose}\ and\ \citenamefont {Taylor}(1983)}]{hose83}%
  \BibitemOpen
  \bibfield  {author} {\bibinfo {author} {\bibfnamefont {G.}~\bibnamefont
  {Hose}}\ and\ \bibinfo {author} {\bibfnamefont {H.~S.}\ \bibnamefont
  {Taylor}},\ }\bibfield  {title} {\enquote {\bibinfo {title} {Quantum
  kolmogorov-arnol'd-moser-like theorem: fundamentals of localization in
  quantum theory},}\ }\href {\doibase 10.1103/PhysRevLett.51.947} {\bibfield
  {journal} {\bibinfo  {journal} {Phys. Rev. Lett.}\ }\textbf {\bibinfo
  {volume} {51}},\ \bibinfo {pages} {947--950} (\bibinfo {year}
  {1983})}\BibitemShut {NoStop}%
\bibitem [{\citenamefont {Chirikov}(1979)}]{chirikov79}%
  \BibitemOpen
  \bibfield  {author} {\bibinfo {author} {\bibfnamefont {B.~V.}\ \bibnamefont
  {Chirikov}},\ }\bibfield  {title} {\enquote {\bibinfo {title} {A universal
  instability of many-dimensional oscillator systems},}\ }\href {\doibase
  https://doi.org/10.1016/0370-1573(79)90023-1} {\bibfield  {journal} {\bibinfo
   {journal} {Physics Reports}\ }\textbf {\bibinfo {volume} {52}},\ \bibinfo
  {pages} {263 -- 379} (\bibinfo {year} {1979})}\BibitemShut {NoStop}%
\bibitem [{\citenamefont {Greene}(1979)}]{greene79}%
  \BibitemOpen
  \bibfield  {author} {\bibinfo {author} {\bibfnamefont {J.~M.}\ \bibnamefont
  {Greene}},\ }\bibfield  {title} {\enquote {\bibinfo {title} {A method for
  determining a stochastic transition},}\ }\href@noop {} {\bibfield  {journal}
  {\bibinfo  {journal} {Journal of Mathematical Physics}\ }\textbf {\bibinfo
  {volume} {20}},\ \bibinfo {pages} {1183--1201} (\bibinfo {year}
  {1979})}\BibitemShut {NoStop}%
\bibitem [{\citenamefont {Husimi}(1940)}]{husimi1940}%
  \BibitemOpen
  \bibfield  {author} {\bibinfo {author} {\bibfnamefont {K.}~\bibnamefont
  {Husimi}},\ }\bibfield  {title} {\enquote {\bibinfo {title} {Some formal
  properties of the density matrix},}\ }\href@noop {} {\bibfield  {journal}
  {\bibinfo  {journal} {Proceedings of the Physico-Mathematical Society of
  Japan. 3rd Series}\ }\textbf {\bibinfo {volume} {22}},\ \bibinfo {pages}
  {264--314} (\bibinfo {year} {1940})}\BibitemShut {NoStop}%
\bibitem [{\citenamefont {Takahashi}\ and\ \citenamefont
  {Sait{\^o}n}(1985)}]{takahashi1985}%
  \BibitemOpen
  \bibfield  {author} {\bibinfo {author} {\bibfnamefont {K.}~\bibnamefont
  {Takahashi}}\ and\ \bibinfo {author} {\bibfnamefont {N.}~\bibnamefont
  {Sait{\^o}n}},\ }\bibfield  {title} {\enquote {\bibinfo {title} {Chaos and
  husimi distribution function in quantum mechanics},}\ }\href@noop {}
  {\bibfield  {journal} {\bibinfo  {journal} {Physical Review Letters}\
  }\textbf {\bibinfo {volume} {55}},\ \bibinfo {pages} {645} (\bibinfo {year}
  {1985})}\BibitemShut {NoStop}%
\bibitem [{\citenamefont {Mackay}\ \emph {et~al.}(1984)\citenamefont {Mackay},
  \citenamefont {Meiss},\ and\ \citenamefont {Percival}}]{mackay84}%
  \BibitemOpen
  \bibfield  {author} {\bibinfo {author} {\bibfnamefont {R.S.}\ \bibnamefont
  {Mackay}}, \bibinfo {author} {\bibfnamefont {J.D.}\ \bibnamefont {Meiss}}, \
  and\ \bibinfo {author} {\bibfnamefont {I.C.}\ \bibnamefont {Percival}},\
  }\bibfield  {title} {\enquote {\bibinfo {title} {Transport in hamiltonian
  systems},}\ }\href {\doibase https://doi.org/10.1016/0167-2789(84)90270-7}
  {\bibfield  {journal} {\bibinfo  {journal} {Physica D: Nonlinear Phenomena}\
  }\textbf {\bibinfo {volume} {13}},\ \bibinfo {pages} {55 -- 81} (\bibinfo
  {year} {1984})}\BibitemShut {NoStop}%
\bibitem [{\citenamefont {Bensimon}\ and\ \citenamefont
  {Kadanoff}(1984)}]{bensimon84}%
  \BibitemOpen
  \bibfield  {author} {\bibinfo {author} {\bibfnamefont {D.}~\bibnamefont
  {Bensimon}}\ and\ \bibinfo {author} {\bibfnamefont {L.~P.}\ \bibnamefont
  {Kadanoff}},\ }\bibfield  {title} {\enquote {\bibinfo {title} {Extended chaos
  and disappearance of kam trajectories},}\ }\href {\doibase
  https://doi.org/10.1016/0167-2789(84)90271-9} {\bibfield  {journal} {\bibinfo
   {journal} {Physica D: Nonlinear Phenomena}\ }\textbf {\bibinfo {volume}
  {13}},\ \bibinfo {pages} {82 -- 89} (\bibinfo {year} {1984})}\BibitemShut
  {NoStop}%
\end{thebibliography}%

\end{document}